\documentclass[twocolumn,pra,showpacs,nofootinbib]{revtex4}

\usepackage[latin1]{inputenc}
\usepackage{amsmath,amssymb,bbm,epsfig}

\newcommand{\assign}{:=}
\newcommand{\bigintlim}{\int}
\newcommand{\bignone}{\,}

\newcommand{\mathd}{\mathrm{d}}
\newcommand{\mathe}{\mathrm{e}}
\newcommand{\mathi}{\mathrm{i}}
\newcommand{\tmem}[1]{{\em #1\/}}
\newcommand{\tmmathbf}[1]{\ensuremath{\boldsymbol{#1}}}
\newcommand{\tmop}[1]{\ensuremath{\operatorname{#1}}}
\newcommand{\um}{-}

\begin{document}
\author{Klaus Hornberger}
\affiliation{Arnold Sommerfeld Center for Theoretical Physics,
Ludwig-Maximilians-Universität München, Theresienstraße 37, 80333 Munich, Germany}
\homepage{www.klaus-hornberger.de}

\author{Bassano Vacchini}
\affiliation{Dipartimento di Fisica dell'Università di Milano and INFN Sezione di Milano, Via Celoria 16, 20133 Milano, Italy}
\date{\today}
\pacs{03.65.Yz, 05.20.Dd, 03.75.-b, 47.45.Ab \hfill Published in \sf Phys. Rev. A 77, 022112 (2008)   }


\title{Monitoring derivation of the quantum linear Boltzmann
equation}

\begin{abstract}
  We show how the effective equation of motion for a distinguished quantum
  particle in an ideal gas environment can be obtained by means of the
  monitoring approach introduced in [EPL 77, 50007 (2007)]. The resulting
  Lindblad master equation accounts for the quantum effects of the scattering
  dynamics in a non-perturbative fashion and it describes decoherence and
  dissipation in a unified framework. It incorporates various established
  equations as limiting cases and reduces to the classical linear Boltzmann
  equation once the state is diagonal in momentum.
\end{abstract}
\maketitle

\section{Introduction}

A basic problem in the field of open quantum dynamics is the question how the
motion of a tracer particle, such as a Brownian particle, is affected by the
presence of a background gas {\cite{Spohn1980a}}. More specifically, one may
consider a single distinguished test particle which moves in the absence of
external forces, but is interacting with an ideal, non-degenerate, and
stationary gas. The elastic collisions with the gas particles will affect the
motional state of the tracer particle, and we are interested in the
appropriate effective equation of motion for its (reduced) density operator
which incorporates the interaction process in a non-perturbative manner. This
master equation is necessarily linear, since it pertains to a single particle,
and it is aptly called, in analogy to the case of a classical tracer particle
{\cite{Cercignani1975a}}, the {\tmem{quantum linear Boltzmann equation}}
(QLBE). However, one should not confuse it with a {\tmem{linearized}} quantum
equation for the single particle gas state of a self-interacting quantum gas,
sometimes called by the same name (though the notation ``linearized quantum
Boltzmann equation'' would seem more fitting).

The dynamics to be described by the QLBE can be quite involved because the
tracer particle may be in a very non-trivial motional state, characterized for
example by the non-classical correlations between different position and
momentum components found in a matter wave interferometer {\cite{Arndt2005a}}.
On the long run, the tracer particle will approach a stationary
``thermalized'' state, while the ever increasing entanglement with the gas
will reduce its quantum coherences already on much shorter time scales. A
limiting case occurs if the tracer particle can be taken as infinitely
massive, so that energy exchange during the collisions can be safely
neglected. In this case one expects pure collisional decoherence, i.e., a
spatial ``localization'' of an extended coherent matter wave into a mixture
with reduced spatial coherence.

This problem was first investigated by Joos and Zeh in a linearized
description {\cite{Joos1985a}}. However, a {\tmem{non-perturbative}} treatment
is required to describe how the spatial coherences in an interfering state get
reduced the more the better the scattered gas particles can ``resolve'' the
different interference paths, and to account for the saturation of this effect
with increasing path difference {\cite{Gallis1990a,Hornberger2003b}}. This
loss of coherence, which may be related to the ``which path'' information
revealed to the environment, was observed experimentally with interfering
fullerene molecules in good quantitative agreement with decoherence theory
{\cite{Hornberger2003a}}.

The situation is much more involved if the ratio $m / M$ between the mass $m$
of the gas particles and the mass $M$ of the tracer particle cannot be
neglected. In this case the particle experiences friction, it will dissipate
its energy and finally thermalize. The appropriate effective equation must
then be able to describe the full interplay of decohering and dissipative
dynamics. An important advancement in this direction was the proposal by
Di\'osi {\cite{Diosi1995a}} of an equation based on a combination of
scattering theory and heuristic arguments. In this derivation a number of
ad-hoc approximations had to be introduced when incorporating the Markov
assumption in order to end up with a time-local master equation in Lindblad
form. As is notorious in non-perturbative derivations of Markovian master
equations, these approximations are not unambiguous and very hard to motivate
microscopically.

One way to overcome this ambiguity problem was recently proposed by one of us
{\cite{Hornberger2007b}}. This method, called the monitoring approach, treats
the Markov assumption not as an approximation to be performed when tracing out
the environmental degrees of freedom, but incorporates it before this trace is
done by combining concepts from the theory of generalized and continuous
measurements with time dependent scattering theory. When applied to the
present case, the essential premise of this approach is to assume that both
the rate and the effect of individual collisions between the tracer and the
gas particle are separately well defined. The Markov assumption then enters by
saying that three-particle collisions are sufficiently unlikely to be safely
neglected, as are subsequent collisions with the same gas molecule within the
relevant time scale. This assumption excludes liquefied or strongly
self-interacting ``gas'' environments, but it seems natural in the case of an
ideal gas in a stationary state. The only real freedom in this framework of
the monitoring approach lies in the choice of two microscopic operators.
Selecting the operators suggested by microscopic scattering theory then leads
to the equation in an unambiguous way.

The present result was already announced in {\cite{Hornberger2006b}}. Here we
give a more detailed derivation{\footnote{We emphasize that the word
``derivation'' is used here in the physicist's sense, implying that arguments
and approximations are invoked which--though physically stringent and leading
to a uniquely distinguished equation--may be hard to substantiate in a proper
mathematical framework. We certainly do not claim to provide a mathematically
rigorous proof, noting that even the classical Boltzmann equation still lacks
such a mathematical derivation.}}, presenting two independent ways of
evaluating the environmental trace. We will also point out that various limits
reduce the QLBE to well-established results. In particular, one obtains the
weak-coupling version of the QLBE, proposed earlier by one of us
{\cite{Vacchini2000a,Vacchini2001b,Vacchini2002a}}, if the appropriate limit
is taken by replacing the scattering amplitudes with their Born approximation.
Other limits lead to the generalized form of the Caldeira-Leggett master
equation {\cite{Caldeira1983a}}, the master equation of pure collisional
decoherence, and the classical linear Boltzmann equation.

The structure of the article is as follows. In Sect.~\ref{sec:monitoring} we
briefly review the monitoring approach and specify the microscopic operators
for the problem at hand. Before delving into the calculations we present the
form of the resulting QLBE in momentum representation in
Sect.~\ref{sec:momemtumQLBE}. This allows us to discuss the relation of the
QLBE to the classical linear Boltzmann equation. Section~\ref{sec:momentum}
then starts out with the calculation in momentum basis and explains why a
straightforward evaluation of the trace is impossible. A first remedy, based
on the restriction to wave packet states of incoming type is given in
Sect.~\ref{sec:wp}. Section \ref{sec:rr} provides an alternative way of doing
the environmental trace, which is based on a formal redefinition of the
scattering operator. Section \ref{sec:refrac} is devoted to calculating the
coherent modification part of the master equation using the same wave packet
technique as in Sect.~\ref{sec:wp}. The basis independent ``operator form'' of
the QLBE is obtained in Sect.~\ref{sec:or}; it shows immediately that the
master equation provides the generator of a completely positive and
translationally invariant quantum dynamical semigroup. Section~\ref{sec:lf}
summarizes the various limiting forms of the QLBE, and we present our
conclusions in Sect.~\ref{sec:cc}.

\begin{flushleft}
  \section{The monitoring master equation for a tracer particle in a
  gas}\label{sec:monitoring}
\end{flushleft}

\subsection{The monitoring master equation}

Let us start with a brief review of the monitoring approach
{\cite{Hornberger2007b}}. It yields a Markovian master equation that is
specified, apart form the system Hamiltonian $\mathsf{H}$, in terms of two
operators, a {\tmem{rate operator}} $\mathsf{\Gamma}$ and a {\tmem{scattering
operator}} $\mathsf{S}$.

The operator $\mathsf{\Gamma}$ is {\tmem{positive}} and in the present
context it has the defining property that its expectation value yields the
probability of collision with the gas particles in the small time interval
$\Delta t$,
\begin{eqnarray}
  \Pr \left( \text{C}_{\Delta t} | \rho \otimes \rho_{\tmop{gas}} \right) & =
  & \tmop{Tr} \left( \text{$\mathsf{\Gamma}$} \left[ \rho \otimes
  \rho_{\tmop{gas}} \right] \right) \Delta t + \mathcal{O} \left( \Delta t^2
  \right) .  \label{eq:probcol}
\end{eqnarray}
Here, $\rho$ is the system density operator which describes, in the present
application, the motional state of the tracer particle. The operator
$\rho_{\tmop{gas}}$ is the effective single particle state of the gas
environment, and it is assumed to be stationary (but not necessarily in
thermal equilibrium). Thus, $\mathsf{\Gamma}$ acts in a two-particle Hilbert
space, and its task is to incorporate the tracer state-dependence of the
collision probability into the dynamic formulation.

The scattering operator $\mathsf{S}$, on the other hand, is {\tmem{unitary}},
and by definition it yields the two particle state after a single collision,
so that, upon tracing over the gas particle, we obtain the new tracer particle
state (in interaction picture) after a single scattering event took place
{\cite{Stenholm1993a}},
\begin{eqnarray}
  \rho' & = & \tmop{Tr}_{\tmop{gas}} \left( \mathsf{S} \left[ \rho \otimes
  \rho_{\tmop{gas}} \right] \mathsf{S}^{\dag} \right) .  \label{eq:rhoprime}
\end{eqnarray}
The monitoring approach {\cite{Hornberger2007b}} now implements the Markov
assumption by combining the state dependence of the collision probability
(\ref{eq:probcol}) with the transformation (\ref{eq:rhoprime}) in a way which
is consistent with the state transformation rules of quantum mechanics
{\cite{Kraus1983a}}, using concepts of the theory of generalized and
continuous measurements {\cite{Busch1991a,Holevo2001a,Jacobs2006a}}. In the
Schr\"odinger picture one thus obtains the effective equation of motion
\begin{eqnarray}
  \frac{\mathd}{\mathd t} \rho & = & \frac{1}{\mathi \hbar} \left[ \mathsf{H},
  \rho \right] + \mathcal{L} \rho + \mathcal{R} \rho .  \label{eq:me0}
\end{eqnarray}
The superoperators $\mathcal{L}$ and $\mathcal{R}$ are best specified in terms
of the nontrivial part $\mathsf{T}$ of the scattering operator $\mathsf{S} =
\mathsf{I} + i \mathsf{T}$. The part $\mathcal{L} \rho$ then takes the form
{\cite{Hornberger2007b}}
\begin{eqnarray}
  \mathcal{L} \rho & = & \tmop{Tr}_{\tmop{gas}} \left( \mathsf{T}
  \mathsf{\Gamma}^{1 / 2} \left[ \rho \otimes \rho_{\tmop{gas}} \right]
  \mathsf{\Gamma}^{1 / 2} \mathsf{T}^{\dag} \right) \nonumber\\
  &  & - \frac{1}{2} \tmop{Tr}_{\tmop{gas}} \left( \mathsf{\Gamma}^{1 / 2}
  \mathsf{T}^{\dag} \mathsf{T}  \mathsf{\Gamma}^{1 / 2} \left[ \rho \otimes
  \rho_{\tmop{gas}} \right] \right) \nonumber\\
  &  & - \frac{1}{2} \tmop{Tr}_{\tmop{gas}} \left( \left[ \rho \otimes
  \rho_{\tmop{gas}} \right]  \mathsf{\Gamma}^{1 / 2} \mathsf{T}^{\dag}
  \mathsf{T} \mathsf{\Gamma}^{1 / 2} \right) .  \label{eq:me1}
\end{eqnarray}
It describes the incoherent evolution of $\rho$ due to the presence of the gas
environment. The part $\mathcal{R} \rho$, on the other hand, is given
by{\footnote{A marginally different expression was given in
Ref.~{\cite{Hornberger2007b}}, see the discussion in Sect.~\ref{sec:refrac}.}}
\begin{eqnarray}
  \mathcal{R} \rho & = & \mathi \tmop{Tr}_{\tmop{gas}} \left( \left[
  \mathsf{\Gamma}^{1 / 2} \tmop{Re} \left( \mathsf{T} \right) 
  \mathsf{\Gamma}^{1 / 2}, \rho \otimes \rho_{\tmop{gas}} \right] \right), 
  \label{eq:Rcal}
\end{eqnarray}
where $\tmop{Re} \left( \mathsf{T} \right) = \left( \mathsf{T} +
\mathsf{T}^{\dag} \right) / 2$. It is responsible for a unitary modification
of the evolution, a renormalization of the system energy due to the coupling
with the environment.

We would like to emphasize that the evolution described by (\ref{eq:me0}) is
non-perturbative in the sense that the collisional transformation described by
$\mathsf{S}$ is not assumed to be weak. Moreover, note that the incoherent
part (\ref{eq:me1}) is manifestly Markovian even before the environmental
trace is done.

\subsection{Rate and scattering operators}\label{sec:operators}

In the framework of the monitoring approach the only essential freedom lies in
the choice of the operators $\mathsf{\Gamma}$, $\mathsf{T}$, and $\mathsf{H}$
appearing in Eqs.~(\ref{eq:me0})-(\ref{eq:Rcal}). In this section we will
specify them on a microscopic basis. Before that it is helpful to consider
with some care $\rho_{\tmop{gas}}$, the effective single particle state of an
ideal gas with number density $n_{\tmop{gas}}$.

To be describable by a normalizable state, the gas must be confined, say with
periodic boundary conditions, to a finite spatial region $\Omega$ with (large)
normalization volume $\left| \Omega \right|$. Let us denote the projector to
this spatial region as
\begin{eqnarray}
  \mathsf{I}_{\Omega} & = & \int_{\Omega} \mathd \tmmathbf{x} \bignone
  |\tmmathbf{x} \rangle \langle \tmmathbf{x}|.  \label{eq:Omegaprojection}
\end{eqnarray}
Using the double-bracket notation $||\tmmathbf{p} \rangle \rangle$ for the
volume-normalized momentum states, the density operators corresponding to
these proper vectors take the form
\begin{eqnarray}
  \rho_{\tmmathbf{p}} & = & ||\tmmathbf{p}_{} \rangle \rangle \langle \langle
  \tmmathbf{p}|| \hspace{0.6em} = \hspace{0.6em} \frac{\left( 2 \pi \hbar
  \right)^3}{| \Omega |} \mathsf{I}_{\Omega} |\tmmathbf{p} \rangle \langle
  \tmmathbf{p}| \mathsf{I}_{\Omega} .  \label{eq:rhopure}
\end{eqnarray}
Here the $|\tmmathbf{p} \rangle$ are the usual improper momentum eigenvectors,
$\langle \tmmathbf{x}|\tmmathbf{p} \rangle = \left( 2 \pi \hbar \right)^{- 3 /
2} {\exp \left( i\tmmathbf{x} \cdot \tmmathbf{p}/ \hbar \right)}$. Since
$\rho_{\tmop{gas}}$ is stationary it must be a convex combination of the pure
momentum states (\ref{eq:rhopure}). It is completely characterized by the gas
momentum distribution $\mu \left( \tmmathbf{p} \right)$, a positive function
satisfying$\int \mathd \tmmathbf{p} \mu \left( \tmmathbf{p} \right) = 1
\bignone$. Thus $\rho_{\tmop{gas}}$ has the form
\begin{eqnarray}
  \text{$\rho _{\tmop{gas}}$} & = & \int \mathd \tmmathbf{p} \, \mu \left(
  \tmmathbf{p} \right) \bignone \rho_{\tmmathbf{p}} \hspace{0.6em} =
  \hspace{0.6em} \frac{\left( 2 \pi \hbar \right)^3}{| \Omega |} 
  \mathsf{I}_{\Omega} \mu ( \mathsf{p}) \mathsf{I}_{\Omega}, 
  \label{eq:rhogasmu}
\end{eqnarray}
where $\mathsf{p}$ is the unrestricted momentum operator of a single gas
particle. This state is normalized, $\tmop{Tr} \left( \text{$\rho
_{\tmop{gas}}$} \right) = 1$, and it is uniform in position, $\langle
\tmmathbf{x}| \rho_{\tmop{gas}} |\tmmathbf{x} \rangle = 1 / | \Omega |$ for
$\tmmathbf{x} \in \Omega$. The most natural choice for $\mu$ is of course the
Maxwell-Boltzmann distribution, see (\ref{eq:muMB}) below, but we will keep
$\mu$ unspecified in order to indicate that the particular form of $\mu$ is
not relevant for most of what follows.

In principle, projections similar to the $\mathsf{I}_{\Omega}$ in
(\ref{eq:rhogasmu}) are also needed when defining the operators
$\mathsf{\Gamma}$, $\mathsf{T}$, and $\mathsf{H}$ of
(\ref{eq:me0})-(\ref{eq:Rcal}). To avoid clumsy notation we will instead
present them in their unrestricted form and take care of the restrictions
during the calculations below.

Since the tracer particle is supposed to move in the absence of external
forces the Hamiltonian part of (\ref{eq:me0}) is given by $\mathsf{H} =
\mathsf{P}^2 / 2 M$, where $\mathsf{P}$ is the momentum operator of the tracer
particle. The two-particle operators $\mathsf{\Gamma}$ and $\mathsf{T}$ depend
on the relative coordinates between tracer and gas particle, and it will be
convenient to denote relative momenta by
\begin{eqnarray}
  \tmop{rel} \left( \tmmathbf{p}, \tmmathbf{P} \right) & \assign &
  \frac{m_{\ast}}{m} \tmmathbf{p}- \frac{m_{\ast}}{M} \tmmathbf{P} 
  \label{eq:reldef}
\end{eqnarray}
with $m_{\ast} = mM / \left( M + m \right)$ the reduced mass. Thus, the
momentum dyadics corresponding to the different factorizations of the total
Hilbert space $\mathcal{H}_{\tmop{tot}} = \mathcal{H} \otimes
\mathcal{H}_{\tmop{gas}} = \mathcal{H}_{\tmop{cm}} \otimes
\mathcal{H}_{\tmop{rel}}$ are related by
\begin{eqnarray}
  |\tmmathbf{P} \rangle \langle \tmmathbf{P}' | \otimes |\tmmathbf{p} \rangle
  \langle \tmmathbf{p}' |_{\tmop{gas}} & = & |\tmmathbf{P}+\tmmathbf{p}
  \rangle \langle \tmmathbf{P}' +\tmmathbf{p}' |_{\tmop{cm}} 
  \label{eq:ptrafo}\\
  &  & \otimes | \tmop{rel} \left( \tmmathbf{p}, \tmmathbf{P} \right) \rangle
  \langle \tmop{rel} \left( \tmmathbf{p}', \tmmathbf{P}' \right)
  |_{\tmop{rel}} . \nonumber
\end{eqnarray}

In classical mechanics, the collision rate is obtained by multiplying the
current density $j_0 \left( \tmmathbf{p}, \tmmathbf{P} \right) =
n_{\tmop{gas}}  \left| \tmop{rel} \left( \tmmathbf{p}, \tmmathbf{P} \right)
\right| / m_{\ast}$ of the relative motion with the total scattering cross
section $\sigma_{\tmop{tot}}$ (which also depends on the relative momentum).
It seems therefore natural to define $\mathsf{\Gamma}$ as the corresponding
operator on $\mathcal{H}_{\tmop{cm}} \otimes \mathcal{H}_{\tmop{rel}}$,
\begin{eqnarray}
  \mathsf{\Gamma} & = & \mathsf{I}_{\tmop{cm}} \otimes \left[ j_0 \left(
  \mathsf{p}, \mathsf{P} \right) \sigma_{\tmop{tot}} \left( \tmop{rel} \left(
  \mathsf{p}, \mathsf{P} \right) \right) \right]_{\tmop{rel}} 
  \label{eq:Gam3}\\
  & = & \bignone \bignone \mathsf{I}_{\tmop{cm}} \otimes \left[
  \mathsf{\Gamma}_0 \right]_{\tmop{rel}} \nonumber
\end{eqnarray}
with
\begin{eqnarray}
  \mathsf{\Gamma}_0 & = & \frac{n_{\tmop{gas}}}{m_{\ast}} \left| \tmop{rel}
  \left( \mathsf{p}, \mathsf{P} \right) \right| \sigma_{\tmop{tot}} \left(
  \tmop{rel} \left( \mathsf{p}, \mathsf{P} \right) \right) . 
  \label{eq:Gamrel}
\end{eqnarray}
Indeed, for normalized and separable particle-gas states the expectation value
of this operator yields the collision rate experienced by the tracer particle,
provided their relative state is of {\tmem{incoming type}}. If the
two-particle state is of outgoing type, on the other hand, the motion of the
relative coordinate is directed away from the origin, so that the particle and
the gas molecule never interact. Still, the operator (\ref{eq:Gam3}) would
yield a finite expectation value in that case, since it depends only on the
modulus of the relative velocity and not its orientation. A proper definition
of $\mathsf{\Gamma}$ should therefore also include a projection to the
subspace of truly incoming relative motional states. Unfortunately, it is
rather difficult to formulate this projection in a way so that one can work
with it in concrete calculations. Therefore, instead of using a more refined
definition we shall stick with Eq.~(\ref{eq:Gam3}) keeping in mind that it is
valid only for incoming states of the relative motion.

As the last step, we have to define the operator $\mathsf{S} = \mathsf{I} + i
\mathsf{T}$ describing the effect of a single collision. It is natural to use
scattering theory for a microscopic definition {\cite{Stenholm1993a}}. The
center-of-mass coordinate then remains unaffected,
\begin{eqnarray}
  \mathsf{S} & = & \mathsf{I}_{\tmop{cm}} \otimes \left[ \mathsf{S}_0
  \right]_{\tmop{rel}}  \label{eq:Smatrix}
\end{eqnarray}
and the scattering operator of the relative coordinates $\mathsf{S}_0 =
\mathsf{I} + i \mathsf{T}_0$ is fully specified in terms of the complex
scattering amplitudes ${f \left( \tmmathbf{p}_f, \tmmathbf{p}_i \right)}$,
which are determined by the inter-particle potential {\cite{Taylor1972a}},
\begin{eqnarray}
  \langle \tmmathbf{p}_f | \mathsf{T}_0 |\tmmathbf{p}_i \rangle & = &
  \frac{1}{2 \pi \hbar} \delta \left( \frac{p_f^2 - p_i^2}{2} \right) f \left(
  \tmmathbf{p}_f, \tmmathbf{p}_i \right) .  \label{eq:Tme}
\end{eqnarray}
Note that a delta-function in (\ref{eq:Tme}) ensures that the energy is
conserved during an elastic collision changing the relative momentum from
$\tmmathbf{p}_i$ to $\tmmathbf{p}_f$.

The scattering amplitude also defines the cross section required in
(\ref{eq:Gam3}). The differential cross section is given by
\begin{eqnarray}
  \sigma \left( \tmmathbf{p}_f, \tmmathbf{p}_i \right) & = & \left| f \left(
  \tmmathbf{p}_f, \tmmathbf{p}_i \right) \right|^2  \label{eq:dcross}
\end{eqnarray}
and the total cross section reads
\begin{eqnarray}
  \sigma_{\tmop{tot}} \left( \tmmathbf{p}_i \right) & = & \int \mathd
  \tmmathbf{n} \bignone \left| f \left( p_i \tmmathbf{n}, \tmmathbf{p}_i
  \right) \right|^2,  \label{eq:totcross}
\end{eqnarray}
where $\tmmathbf{n}$ is a unit vector with $\mathd \tmmathbf{n}$ the
associated solid angle element. 
\begin{figure*}[tb]
  \resizebox{15cm}{!}{\epsfig{file=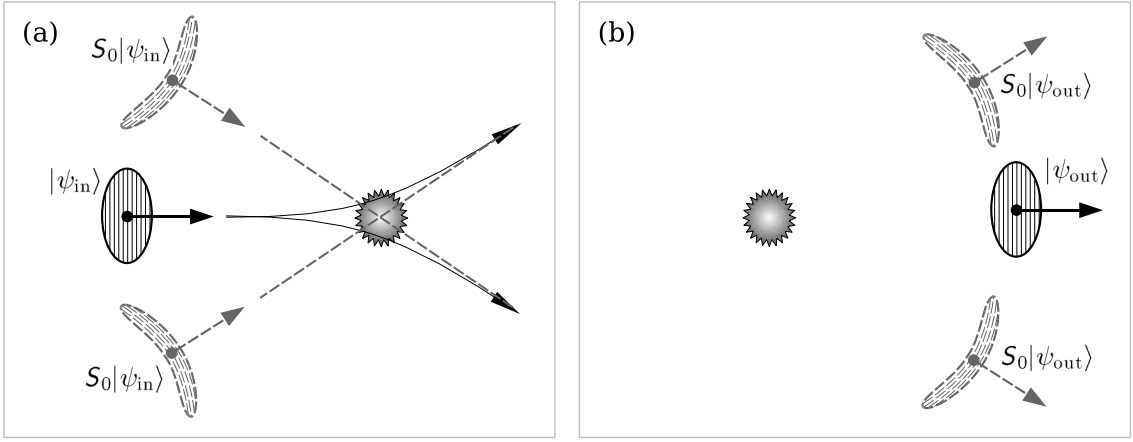}}
  \caption{\label{fig:cartoon}Action of the S-matrix when applied to localized
  wave packets of the incoming and the outgoing type. (a) An incoming wave
  packet $| \psi_{\tmop{in}} \rangle$ is transformed in such a way that the
  free motion of the resulting state $\mathsf{S}_0 | \psi_{\tmop{in}} \rangle$
  (indicated by the dashed curves) converges with the dynamically scattered
  wave packet at large times {\cite{Taylor1972a}}. (b) An outgoing wave
  packet, whose forward time evolution will be unaffected by the scattering
  potential, gets strongly transformed by $\mathsf{S}_0$. This is due to the
  inverse time evolution involved in the definition of the the S-matrix. To
  prevent this unwanted transformation one should $\left( i \right)$ either
  attribute a vanishing collision rate to all outgoing states or {\tmem{(ii)}}
  modify the $\mathsf{S}_0$ operator such that it leaves all outgoing wave
  packets unaffected. Evaluations of $\mathcal{L} \rho$ based on these two
  strategies are given in Sect.~\ref{sec:wp} and Sect.~\ref{sec:rr},
  respectively.}
\end{figure*}

It is important to keep in mind that the S-matrix (\ref{eq:Smatrix}) provided
by scattering theory is physically meaningful only for proper incoming states
of the relative motion, even though it is defined on the whole Hilbert space.
This is schematically shown in Fig.~\ref{fig:cartoon} where we contrast the
action of $\mathsf{S}_0$ on an incoming wave packet with its effect on an
outgoing state, showing that an outgoing wave packet may get spuriously
transformed. The reason is that the Møller operator $\hat{\Omega}_+ = \lim_{t
\rightarrow \infty} \mathsf{U} \left( t \right) \mathsf{U}_0 \left( - t
\right)$ used to construct $\mathsf{S}_0 = \hat{\Omega}^{\dag}_-
\hat{\Omega}_+$ involves a free {\tmem{backward}} evolution in time
$\mathsf{U}_0 \left( \um t \right)$, followed by a forward motion $\mathsf{U}
\left( t \right)$ in the presence of the interaction potential. To avoid this
undesired transformation in (\ref{eq:me0}) we must either ensure that outgoing
wave packets contribute with a zero collision rate or we have to modify
$\mathsf{S}_0$ such that it leaves the outgoing contributions invariant.

\section{The QLBE in momentum representation}\label{sec:momemtumQLBE}

Before we proceed to derive the quantum linear Boltzmann equation let us
present the result in the basis of improper momentum eigenstates
$|\tmmathbf{P} \rangle$, where it takes a particularly simple form. This
permits to introduce the complex rate function $M_{\tmop{in}}$ to be evaluated
in Sects.~\ref{sec:momentum}--\ref{sec:rr}, and to discuss its relation to the
classical rate densities of the collision kernel.

We will see that the momentum representation of the incoherent part
(\ref{eq:me1}) of the QLBE can be written in terms of a single complex
function, and that it takes the form
\begin{eqnarray}
  &  & \langle \tmmathbf{P}| \mathcal{L} \rho |\tmmathbf{P}' \rangle
  \nonumber\\
  &  & = \int \mathd \tmmathbf{Q} \bignone \langle \tmmathbf{P}-\tmmathbf{Q}|
  \rho |\tmmathbf{P}' -\tmmathbf{Q} \rangle M_{\tmop{in}} \left( \tmmathbf{P},
  \tmmathbf{P}' ; \tmmathbf{Q} \right)  \label{eq:Dtrho1}\\
  &  & \phantom{=} - \frac{\langle \tmmathbf{P}| \rho |\tmmathbf{P}'
  \rangle}{2} \int \mathd \tmmathbf{Q} \, M_{\tmop{in}} \left(
  \tmmathbf{P}+\tmmathbf{Q}, \tmmathbf{P}+\tmmathbf{Q}; \tmmathbf{Q} \right)
  \nonumber\\
  &  & \phantom{=} - \frac{\langle \tmmathbf{P}| \rho |\tmmathbf{P}'
  \rangle}{2} \int \mathd \tmmathbf{Q} \, M_{\tmop{in}} \left( \tmmathbf{P}'
  +\tmmathbf{Q}, \tmmathbf{P}' +\tmmathbf{Q}; \tmmathbf{Q} \right) . \nonumber
\end{eqnarray}
The function $M_{\tmop{in}} \left( \tmmathbf{P}, \tmmathbf{P}' ; \tmmathbf{Q}
\right)$ is defined below in Eq.~(\ref{eq:Mexp}) [see also (\ref{eq:Min3}) and
(\ref{eq:Minrel})]. In order to highlight the relation to the classical linear
Boltzmann equation, let us first note that the master equation takes the
shorter form
\begin{eqnarray}
  \langle \tmmathbf{P}| \mathcal{L} \rho |\tmmathbf{P}' \rangle & = & \int
  \mathd \tmmathbf{Q} \, M_{\tmop{in}} \left( \tmmathbf{P}, \tmmathbf{P}' ;
  \tmmathbf{Q} \right) \langle \tmmathbf{P}-\tmmathbf{Q}| \rho |\tmmathbf{P}'
  -\tmmathbf{Q} \rangle \nonumber\\
  &  & - \frac{1}{2} \left[ M_{\tmop{out}}^{\tmop{cl}} \left( \tmmathbf{P}
  \right) + M_{\tmop{out}}^{\tmop{cl}} \left( \tmmathbf{P}' \right) \right]
  \langle \tmmathbf{P}| \rho |\tmmathbf{P}' \rangle \nonumber\\
  &  &  \label{eq:Dtrho2}
\end{eqnarray}
once we introduce
\begin{eqnarray}
  M_{\tmop{out}}^{\tmop{cl}} \left( \tmmathbf{P} \right) & \assign &
  \bigintlim \mathd \tmmathbf{Q} \, M_{\tmop{in}} \left(
  \tmmathbf{P}+\tmmathbf{Q}, \tmmathbf{P}+\tmmathbf{Q}; \tmmathbf{Q} \right) .
  \label{eq:Mclout}
\end{eqnarray}
As indicated by its name, this positive function gives the rate known from
the{\tmem{ classical}} linear Boltzmann equation {\cite{Cercignani1975a}} for
a particle with momentum $\tmmathbf{P}$ to be scattered to a different
momentum. It involves an integration over all initial gas momenta
$\tmmathbf{p}_0$ and all momentum exchanges $\tmmathbf{Q}$ subject to the
restriction implied by energy conservation,
\begin{eqnarray}
  M^{\tmop{cl}}_{\tmop{out}} \left( \tmmathbf{P} \right) & = &
  \frac{n_{\tmop{gas}}}{m_{\ast}} \bigintlim \mathd \tmmathbf{p}_0 \mathd
  \tmmathbf{Q} \, \mu \left( \tmmathbf{p}_0 \right) \nonumber\\
  &  & \times \delta \left( \frac{\left| \tmop{rel} \left( \tmmathbf{p}_0,
  \tmmathbf{P} \right) \right|^2 - \left| \tmop{rel} \left( \tmmathbf{p}_0,
  \tmmathbf{P} \right) +\tmmathbf{Q} \right|^2}{2} \right) \nonumber\\
  &  & \times \sigma \left( \tmop{rel} \left( \tmmathbf{p}_0, \tmmathbf{P}
  \right) +\tmmathbf{Q}, \tmop{rel} \left( \tmmathbf{p}_0, \tmmathbf{P}
  \right) \right) .  \label{eq:Mcloutexp}
\end{eqnarray}
Here, $\mu$ is the gas momentum distribution function from
(\ref{eq:rhogasmu}), the function $\tmop{rel} \left( \tmmathbf{p},
\tmmathbf{P} \right)$ is defined in (\ref{eq:reldef}), and $\sigma$ is the
differential cross section (\ref{eq:dcross}). Carrying out the
$\tmmathbf{Q}$-integration one can write the rate in terms of the total cross
section (\ref{eq:totcross}),
\begin{eqnarray}
  M^{\tmop{cl}}_{\tmop{out}} \left( \tmmathbf{P} \right) & = &
  \frac{n_{\tmop{gas}}}{m_{\ast}} \bigintlim \mathd \tmmathbf{p}_0 \mu \left(
  \tmmathbf{p}_0 \right)  \left| \tmop{rel} \left( \tmmathbf{p}_0,
  \tmmathbf{P} \right) \right| \nonumber\\
  &  & \times \sigma_{\tmop{tot}} \left( \tmop{rel} \left( \tmmathbf{p}_0,
  \tmmathbf{P} \right) \right) .  \label{eq:Mcloutexp2}
\end{eqnarray}
It follows that the dynamics described by the ``loss term'' in
(\ref{eq:Dtrho1}), (\ref{eq:Dtrho2}) is fully specified by the rate in the
corresponding classical equation (which involves of course a quantum
mechanical cross section). The term leads to a reduction of the momentum
coherences $\langle \tmmathbf{P}| \rho |\tmmathbf{P}' \neq \tmmathbf{P}
\rangle$ with a rate given by the arithmetic mean of the loss rates of the
corresponding diagonal elements, $\langle \tmmathbf{P}| \rho |\tmmathbf{P}
\rangle$ and $\langle \tmmathbf{P}' | \rho |\tmmathbf{P}' \rangle$, and these
momentum populations, in turn, are depleted in the same way as the momentum
distribution function of the classical linear Boltzmann equation.

The ``gain term'' in (\ref{eq:Dtrho1}), (\ref{eq:Dtrho2}) is also related to
the classical linear Boltzmann equation, but only on the diagonal
$\tmmathbf{P}=\tmmathbf{P}'$. As one expects, $M_{\tmop{in}}$ is positive on
the diagonal, and equal to the rate density $M_{\tmop{in}}^{\tmop{cl}}$ from
the classical linear Boltzmann equation for the tracer particle to end up in
momentum $\tmmathbf{P}_f$ after a momentum gain of $\tmmathbf{Q}$,
\begin{eqnarray}
  &  & M_{\tmop{in}}^{\tmop{cl}} \left( \tmmathbf{P}_f ; \tmmathbf{Q} \right)
  \nonumber\\
  &  & = \hspace{0.6em} M_{\tmop{in}} \left( \tmmathbf{P}_f, \tmmathbf{P}_f ;
  \tmmathbf{Q} \right) \hspace{0.6em} \equiv \hspace{0.6em} M^{\tmop{cl}}
  \left( \tmmathbf{P}_f -\tmmathbf{Q} \rightarrow \tmmathbf{P}_f \right)
  \nonumber\\
  &  & = \hspace{0.6em} \frac{n_{\tmop{gas}}}{m_{\ast}} \bigintlim \mathd
  \tmmathbf{p}_0 \, \mu \left( \tmmathbf{p}_0 \right)  \nonumber\\
  &  & \phantom{= \hspace{0.6em}} \times \delta \left( \frac{\left|
  \tmop{rel} \left( \tmmathbf{p}_0 -\tmmathbf{Q}, \tmmathbf{P}_f \right)
  \right|^2 - \left| \tmop{rel} \left( \tmmathbf{p}_0, \tmmathbf{P}_f
  -\tmmathbf{Q} \right) \right|^2}{2} \right) \nonumber\\
  &  & \phantom{= \hspace{0.6em}} \times \sigma \left( \tmop{rel} \left(
  \tmmathbf{p}_0 -\tmmathbf{Q}, \tmmathbf{P}_f \right), \tmop{rel} \left(
  \tmmathbf{p}_0, \tmmathbf{P}_f -\tmmathbf{Q} \right) \right) . 
  \label{eq:Mclin}
\end{eqnarray}
The second equality in (\ref{eq:Mclin}) introduces the notation
\begin{eqnarray}
  M^{\tmop{cl}} \left( \tmmathbf{P}_i \rightarrow \tmmathbf{P}_f \right) &
  \assign & M_{\tmop{in}}^{\tmop{cl}} \left( \tmmathbf{P}_f ; \tmmathbf{P}_f
  -\tmmathbf{P}_i \right),  \label{eq:Mcl}
\end{eqnarray}
for the rate density corresponding to a change of momentum $\tmmathbf{P}_i$ to
$\tmmathbf{P}_f$. It will be useful for the discussion of the classical linear
Boltzmann equation in Sect.~\ref{sec:lf}, and it yields the classical out rate
(\ref{eq:Mclout}) as $M_{\tmop{out}}^{\tmop{cl}} \left( \tmmathbf{P} \right) =
{\int \mathd \bignone \tmmathbf{P}_f} M^{\tmop{cl}} \left( \tmmathbf{P}_{}
\rightarrow \tmmathbf{P}_f \right)$ thus ensuring the conservation of
probability.

For $\tmmathbf{P} \neq \tmmathbf{P}'$ the function $M_{\tmop{in}}$ is in
general complex-valued, and it has a rather complicated form when stated with
its explicit dependence on $\tmmathbf{P}$, $\tmmathbf{P}'$, and
$\tmmathbf{Q}$: \begin{widetext}
\begin{eqnarray}
  M_{\tmop{in}} \left( \tmmathbf{P}, \tmmathbf{P}' ; \tmmathbf{Q} \right) & =
  & \frac{n_{\tmop{gas}}}{m_{\ast}} \int \mathd \tmmathbf{p}_0 \, \mu^{1 / 2}
  \left( \tmmathbf{p}_0 + \frac{m}{M} \frac{\tmmathbf{P}_{\|}
  -\tmmathbf{P}_{\|}'}{2} \right) \mu^{1 / 2} \left( \tmmathbf{p}_0 -
  \frac{m}{M} \frac{\tmmathbf{P}_{\|} -\tmmathbf{P}_{\|}'}{2} \right)
  \nonumber\\
  &  & \times f \left( \tmop{rel} \left( \tmmathbf{p}_0 -\tmmathbf{Q},
  \tmmathbf{P}- \frac{\tmmathbf{P}_{\|} -\tmmathbf{P}_{\|}'}{2} \right),
  \tmop{rel} \left( \tmmathbf{p}_0, \tmmathbf{P}- \frac{\tmmathbf{P}_{\|}
  -\tmmathbf{P}_{\|}'}{2} -\tmmathbf{Q} \right) \right) \nonumber\\
  &  & \times f^{\ast} \left( \tmop{rel} \left( \tmmathbf{p}_0 -\tmmathbf{Q},
  \tmmathbf{P}' + \frac{\tmmathbf{P}_{\|} -\tmmathbf{P}_{\|}'}{2} \right),
  \tmop{rel} \left( \tmmathbf{p}_0, \tmmathbf{P}' + \frac{\tmmathbf{P}_{\|}
  -\tmmathbf{P}_{\|}'}{2} -\tmmathbf{Q} \right) \right) \nonumber\\
  &  & \times \delta \left( \frac{\tmop{rel} \left( \tmmathbf{p}_0
  -\tmmathbf{Q}, \frac{\tmmathbf{P}+\tmmathbf{P}'}{2} \right)^2 - \tmop{rel}
  \left( \tmmathbf{p}_0, \frac{\tmmathbf{P}+\tmmathbf{P}'}{2} -\tmmathbf{Q}
  \right)^2}{2} \right) .  \label{eq:Mexp}
\end{eqnarray}
\end{widetext} Here we used the abbreviations
\begin{eqnarray}
  \tmmathbf{P}_{\|} & \assign & \frac{\tmmathbf{P} \cdot \tmmathbf{Q}}{Q^2}
  \tmmathbf{Q}  \label{eq:Ppar}
\end{eqnarray}
and
\begin{eqnarray}
  \tmmathbf{P}'_{\|} & \assign & \frac{\tmmathbf{P}' \cdot \tmmathbf{Q}}{Q^2}
  \tmmathbf{Q}  \label{eq:Pparp}
\end{eqnarray}
for the contributions in $\tmmathbf{P}$ and $\tmmathbf{P}'$ parallel to the
momentum exchange $\tmmathbf{Q} \neq 0$. This form of $M_{\tmop{in}} \left(
\tmmathbf{P}, \tmmathbf{P}' ; \tmmathbf{Q} \right)$ clearly reduces to the
diagonal expression (\ref{eq:Mclin}) when $\tmmathbf{P}$ approaches
$\tmmathbf{P}'$. The curious appearance of the $\tmmathbf{P}_{\|}$ and
$\tmmathbf{P}_{\|}'$ contributions in (\ref{eq:Mexp}) ensures, in combination
with the delta-function, that the modulus of the initial and the final
relative momentum are equal in both scattering amplitudes. This will be more
obvious below in Sect.~\ref{sec:momentum}, where suitable relative coordinates
are introduced. The energy conservation is thus manifestly guaranteed for each
of the scattering amplitudes separately, while the arguments will differ in
general.

One of the important properties of the ``complex rate'' (\ref{eq:Mexp}) is
that it admits a factorization of the $\tmmathbf{P}$- and
$\tmmathbf{P}'$-dependence, which will be crucial later on, when we formulate
the master equation in its representation-independent ``operator form''.
Specifically, it will be shown in Sect.~\ref{sec:or} that $M_{\tmop{in}}$ can
be written as a two-dimensional integration over the set $\tmmathbf{Q}^{\bot}
= \left\{ \tmmathbf{p} \in \mathbbm{R}^3 : \tmmathbf{p} \cdot \tmmathbf{Q}= 0
\right\}$ of momenta perpendicular to the momentum exchange $\tmmathbf{Q}$.
This way the integrand factorizes into a product of $\tmmathbf{P}$- and
$\tmmathbf{P}'$-dependent terms,
\begin{eqnarray}
  M_{\tmop{in}} \left( \tmmathbf{P}, \tmmathbf{P}' ; \tmmathbf{Q} \right) & =
  & \int_{\tmmathbf{Q}^{\bot}} \mathd \tmmathbf{p} \, L \left( \tmmathbf{p},
  \tmmathbf{P}-\tmmathbf{Q}; \tmmathbf{Q} \right)  \nonumber\\
  &  & \times L^{\ast} \left( \tmmathbf{p}, \tmmathbf{P}' -\tmmathbf{Q};
  \tmmathbf{Q} \right) .  \label{eq:Min3}
\end{eqnarray}
The functions
\begin{align}  \label{eq:Ldef}
  &   L \left( \tmmathbf{p}, \tmmathbf{P}; \tmmathbf{Q} \right) 
\\
  &   = \hspace{0.6em} \sqrt{\frac{n_{\tmop{gas}} m}{Qm_{\ast}^2 }} \, \mu
  \left( \tmmathbf{p}_{\bot \tmmathbf{Q}} + \left( 1 + \frac{m}{M} \right)
  \frac{\tmmathbf{Q}}{2} + \frac{m}{M} \tmmathbf{P}_{\|\tmmathbf{Q}}
  \right)^{1 / 2} \nonumber\\
  &   \phantom{= \hspace{0.6em}} \times f \left( \tmop{rel} \left(
  \tmmathbf{p}_{\bot \tmmathbf{Q}}, \tmmathbf{P}_{\bot \tmmathbf{Q}} \right) -
  \frac{\tmmathbf{Q}}{2}, \tmop{rel} \left( \tmmathbf{p}_{\bot \tmmathbf{Q}},
  \tmmathbf{P}_{\bot \tmmathbf{Q}} \right) + \frac{\tmmathbf{Q}}{2} \right),
  \nonumber
\end{align}
involve $\tmmathbf{P}_{\|}$ defined in (\ref{eq:Ppar}) and $\tmmathbf{P}_{\bot
\tmmathbf{Q}} \assign \tmmathbf{P}-\tmmathbf{P}_{\|\tmmathbf{Q}}$. In the
representation-independent form of the master equation they turn into
operator-valued expressions, see Sect.~\ref{sec:or}.

Concerning the coherent modification of the QLBE, it will be shown in
Sect.~\ref{sec:refrac} that the momentum representation of the corresponding
term (\ref{eq:Rcal}) reads
\begin{eqnarray}
  \langle \tmmathbf{P}| \mathcal{R} \rho |\tmmathbf{P}' \rangle & \text{} = &
  \frac{E_{\text{n}} \left( \tmmathbf{P} \right) - E_{\text{n}} \left(
  \tmmathbf{P}' \right)}{i \hbar} \langle \tmmathbf{P}| \rho |\tmmathbf{P}'
  \rangle  \label{eq:cohexp1}
\end{eqnarray}
with
\begin{eqnarray}
  E_{\text{n}} \left( \tmmathbf{P} \right) & = & - 2 \pi \hbar^2 
  \frac{n_{\tmop{gas}}}{m_{\ast}} \int \mathd \tmmathbf{p}_0 \, \mu \left(
  \tmmathbf{p}_0 \right) \nonumber\\
  &  & \times \tmop{Re} \left[ f \left( \tmop{rel} \left( \tmmathbf{p}_0,
  \tmmathbf{P} \right), \tmop{rel} \left( \tmmathbf{p}_0, \tmmathbf{P} \right)
  \right) \right] .  \label{eq:cohexp2}
\end{eqnarray}
This shows how the presence of the gas changes the energy of the particle with
respect to the vacuum. This energy shift depends on the particle momentum and
is determined by the real part of the average {\tmem{forward}} scattering
amplitude. This phenomenon is well known in the field of neutron and atom
interference, and it is usually accounted for by introducing an index of
refraction, see Sect.~\ref{sec:ior}.

\section{Evaluation in the momentum basis}\label{sec:momentum}

\subsection{Transformation to relative coordinates}

Our main task in deriving the quantum linear Boltzmann equation is to evaluate
the expressions (\ref{eq:me1}) and (\ref{eq:Rcal}), which is best done in the
momentum representation. Starting with the incoherent part $\mathcal{L} \rho$,
the cyclicity under the trace yields \begin{widetext}
\begin{eqnarray}
  \langle \tmmathbf{P}| \mathcal{L} \rho |\tmmathbf{P}' \rangle & = & \int
  \mathd \tmmathbf{Q} \mathd \tmmathbf{Q}' \, \bignone \langle
  \tmmathbf{P}-\tmmathbf{Q}| \rho |\tmmathbf{P}' -\tmmathbf{Q}' \rangle M
  \left( \tmmathbf{P}, \tmmathbf{P}' ; \tmmathbf{Q}, \tmmathbf{Q}' \right)
  \nonumber\\
  &  & - \frac{1}{2} \int \mathd \tmmathbf{P}_0 \, \langle \tmmathbf{P}_0 |
  \rho |\tmmathbf{P}' \rangle \int \mathd \tmmathbf{P}_f \bignone \, M \left(
  \tmmathbf{P}_f, \tmmathbf{P}_f ; \tmmathbf{P}_f -\tmmathbf{P}_0,
  \tmmathbf{P}_f -\tmmathbf{P} \right) \nonumber\\
  &  & - \frac{1}{2} \int \mathd \tmmathbf{P}_0' \, \langle \tmmathbf{P}|
  \rho |\tmmathbf{P}_0' \rangle \int \mathd \tmmathbf{P}_f \bignone \, M
  \left( \tmmathbf{P}_f, \tmmathbf{P}_f ; \tmmathbf{P}_f -\tmmathbf{P}',
  \tmmathbf{P}_f -\tmmathbf{P}_0' \right) \nonumber
\end{eqnarray}
with
\begin{eqnarray}
  M \left( \tmmathbf{P}, \tmmathbf{P}' ; \tmmathbf{Q}, \tmmathbf{Q}' \right) &
  = & \langle \tmmathbf{P}| \tmop{Tr}_{\tmop{gas}} \left(  \mathsf{T}
  \mathsf{\Gamma}_{}^{1 / 2}  \left[ |\tmmathbf{P}-\tmmathbf{Q} \rangle
  \langle \tmmathbf{P}' -\tmmathbf{Q}' | \otimes \rho_{\tmop{gas}} \right]
  \mathsf{\Gamma}_{}^{1 / 2} \mathsf{T^{\dag}} \right) |\tmmathbf{P}' \rangle
  .  \label{eq:Mdef}
\end{eqnarray}
\end{widetext} Upon inserting the stationary gas state (\ref{eq:rhogasmu}) into
(\ref{eq:Mdef}) we can simplify the expression by transforming from the
two-particle coordinates to the center-of-mass and relative coordinates using
(\ref{eq:ptrafo}). Since $\mathsf{\Gamma}$ and $\mathsf{T}$ depend only on the
relative motion, see (\ref{eq:Gam3}) and (\ref{eq:Smatrix}), one thus finds
\begin{eqnarray}
  &  & M \left( \tmmathbf{P}, \tmmathbf{P}' ; \tmmathbf{Q}, \tmmathbf{Q}'
  \right) \nonumber\\
  &  & = \hspace{0.6em} \delta \left( \tmmathbf{Q}-\tmmathbf{Q}' \right) 
  \frac{\left( 2 \pi \hbar \right)^3}{| \Omega |} \int \mathd \tmmathbf{p}_0
  \, \mu \left( \tmmathbf{p}_0 \right) \nonumber\\
  &  & \phantom{= \hspace{0.6em}} \times \langle \tmop{rel} \left(
  \tmmathbf{p}_0 -\tmmathbf{Q}, \tmmathbf{P} \right) | \mathsf{T}_0
  \mathsf{\Gamma}^{1 / 2}_0 | \tmop{rel} \left( \tmmathbf{p}_0,
  \tmmathbf{P}-\tmmathbf{Q} \right) \rangle \nonumber\\
  &  & \phantom{= \hspace{0.6em}} \times \langle \tmop{rel} \left(
  \tmmathbf{p}_0, \tmmathbf{P}' -\tmmathbf{Q} \right) | \mathsf{\Gamma}_0^{1 /
  2} \mathsf{T^{\dag}_0} | \tmop{rel} \left( \tmmathbf{p}_0 -\tmmathbf{Q},
  \tmmathbf{P}' \right) \rangle_{} \nonumber\\
  &  & = : \hspace{0.6em} \delta \left( \tmmathbf{Q}-\tmmathbf{Q}' \right)
  M_{\tmop{in}} \left( \tmmathbf{P}, \tmmathbf{P}' ; \tmmathbf{Q} \right), 
  \label{eq:Mfull}
\end{eqnarray}
as anticipated above in (\ref{eq:Dtrho1}).

It is now helpful to introduce functions of $\tmmathbf{p}_0$,
\begin{eqnarray}
  \tmmathbf{p}_i & = & \tmop{rel} \left( \tmmathbf{p}_0,
  \frac{\tmmathbf{P}+\tmmathbf{P}'}{2} -\tmmathbf{Q} \right) 
  \label{eq:pidef}\\
  \tmmathbf{p}_f & = & \tmop{rel} \left( \tmmathbf{p}_0 -\tmmathbf{Q},
  \frac{\tmmathbf{P}+\tmmathbf{P}'}{2} \right),  \label{eq:pfdef}
\end{eqnarray}
which denote the {\tmem{mean}} of the pairs of initial and final relative
momenta appearing in (\ref{eq:Mfull}). We also set
\begin{eqnarray}
  \tmmathbf{q} \hspace{0.6em} & = & \tmop{rel} \left( 0,
  \frac{\tmmathbf{P}-\tmmathbf{P}'}{2} \right) .  \label{eq:qdef}
\end{eqnarray}
These definitions imply the relations
\begin{eqnarray}
  \tmmathbf{p}_f +\tmmathbf{q} & = & \tmop{rel} \left( \tmmathbf{p}_0
  -\tmmathbf{Q}, \tmmathbf{P} \right) \nonumber\\
  \tmmathbf{p}_f -\tmmathbf{q} & = & \tmop{rel} \left( \tmmathbf{p}_0
  -\tmmathbf{Q}, \tmmathbf{P}' \right) \nonumber\\
  \tmmathbf{p}_i +\tmmathbf{q} & = & \tmop{rel} \left( \tmmathbf{p}_0,
  \tmmathbf{P}-\tmmathbf{Q} \right) \nonumber\\
  \tmmathbf{p}_i -\tmmathbf{q} & = & \tmop{rel} \left( \tmmathbf{p}_0,
  \tmmathbf{P}' -\tmmathbf{Q} \right) \nonumber\\
  \tmmathbf{p}_i -\tmmathbf{p}_f & = & \tmmathbf{Q}, 
\end{eqnarray}
which are noted here for later reference. Moreover, for given $\tmmathbf{q}$
we shall write
\begin{eqnarray}
  \tmmathbf{q}_{\|} & \equiv & \frac{\tmmathbf{q} \cdot \left( \tmmathbf{p}_f
  -\tmmathbf{p}_i \right)}{\left( \tmmathbf{p}_f -\tmmathbf{p}_i \right)^2}
  \left( \tmmathbf{p}_f -\tmmathbf{p}_i \right)  \label{eq:qpar}\\
  \tmmathbf{q}_{\bot} & \equiv & \tmmathbf{q}-\tmmathbf{q}_{\|} \hspace{0.6em}
  \label{eq:qbot}
\end{eqnarray}
to denote the components parallel and perpendicular to the momentum exchange
$\tmmathbf{Q}= \tmmathbf{p}_i -\tmmathbf{p}_f$.

The complex rate density defined in the second equality of (\ref{eq:Mfull})
now takes the form
\begin{eqnarray}
  M_{\tmop{in}} \left( \tmmathbf{P}, \tmmathbf{P}' ; \tmmathbf{Q} \right) & =
  & \int \mathd \tmmathbf{p}_0 \, \mu \left( \tmmathbf{p}_0 \right) 
  \frac{\left( 2 \pi \hbar \right)^3}{| \Omega |} \nonumber\\
  &  & \times \langle \tmmathbf{p}_f +\tmmathbf{q}| \mathsf{T}_0
  \mathsf{\Gamma}^{1 / 2}_0 |\tmmathbf{p}_i +\tmmathbf{q} \rangle \nonumber\\
  &  & \times \langle \tmmathbf{p}_i -\tmmathbf{q}| \mathsf{\Gamma}_0^{1 / 2}
  \mathsf{T^{\dag}_0} |\tmmathbf{p}_f -\tmmathbf{q} \rangle . \nonumber
\end{eqnarray}
We can write it as the average over the gas momentum distribution function
$\mu$ of a rate density in the center-of-mass frame,
\begin{eqnarray}
  M_{\tmop{in}} \left( \tmmathbf{P}, \tmmathbf{P}' ; \tmmathbf{Q} \right) & =
  & \int \mathd \tmmathbf{p}_0 \, \mu \left( \tmmathbf{p}_0 \right)
  m_{\tmop{in}} \left( \tmmathbf{p}_f, \tmmathbf{p}_i ; \tmmathbf{q} \right), 
  \label{eq:Minrel}
\end{eqnarray}
thus formally introducing
\begin{eqnarray}
  m_{\tmop{in}} \left( \tmmathbf{p}_f, \tmmathbf{p}_i ; \tmmathbf{q} \right) &
  = & \frac{\left( 2 \pi \hbar \right)^3}{| \Omega |} \langle \tmmathbf{p}_f
  +\tmmathbf{q}| \mathsf{T}_0 \mathsf{\Gamma}^{1 / 2}_0 |\tmmathbf{p}_i
  +\tmmathbf{q} \rangle \nonumber\\
  &  & \times \langle \tmmathbf{p}_i -\tmmathbf{q}| \mathsf{\Gamma}_0^{1 / 2}
  \mathsf{T^{\dag}_0} |\tmmathbf{p}_f -\tmmathbf{q} \rangle 
  \label{eq:mreldef}
\end{eqnarray}
in terms of matrix elements of $\mathsf{T}_0 \mathsf{\Gamma}^{1 / 2}_0$ with
respect to the relative momentum coordinates. This expression should be viewed
here as a generalized function in the sense of distributions, with independent
variables $\tmmathbf{p}_f, \tmmathbf{p}_i$, and $\tmmathbf{q}$.

The main aim of the following sections is to show, in two independent lines of
argument, that the expression (\ref{eq:mreldef}) should be understood as
\begin{align}
  m_{\tmop{in}} \left( \tmmathbf{p}_f, \tmmathbf{p}_i ; \tmmathbf{q} \right) 
  = & \left\{ \begin{array}{ll}
    m_{\tmop{in}} \left( \tmmathbf{p}_f, \tmmathbf{p}_i ; \tmmathbf{q}_{\bot}
    \right) & \text{if $\tmmathbf{q} \cdot \left( \tmmathbf{p}_f
    -\tmmathbf{p}_i \right) = 0$}\\
    0 & \text{otherwise},
  \end{array} \right. \nonumber\\
    &  \label{eq:mclaim0}
\end{align}
with
\begin{align}
  m_{\tmop{in}} \left( \tmmathbf{p}_f, \tmmathbf{p}_i ; \tmmathbf{q}_{\bot}
  \right)  = & \frac{n_{\tmop{gas}}}{m_{\ast}} \delta \left(
  \frac{\tmmathbf{p}_f^2 -\tmmathbf{p}_i^2}{2} \right) f \left( \tmmathbf{p}_f
  +\tmmathbf{q}_{\bot}, \tmmathbf{p}_i +\tmmathbf{q}_{\bot} \right)
  \nonumber\\
    & \times f^{\ast} \left( \tmmathbf{p}_f -\tmmathbf{q}_{\bot},
  \tmmathbf{p}_i -\tmmathbf{q}_{\bot} \right) .  \label{eq:mclaim}
\end{align}
Note that this term involves a single delta-function and the abbreviation
$\tmmathbf{q}_{\bot}$ just defined in (\ref{eq:qbot}).

\subsection{Diagonal evaluation of the trace}\label{sec:diagev}

As a first step, let us try to evaluate (\ref{eq:mreldef}) in a
straightforward fashion by maintaining that the operator $\mathsf{\Gamma}_0$
is diagonal in the relative momentum coordinates. Equation~(\ref{eq:Gamrel})
then implies
\begin{eqnarray}
  \mathsf{\Gamma}^{1 / 2}_0 |\tmmathbf{p} \rangle & = & \sqrt{\Gamma_0 \left(
  \tmmathbf{p} \right)} |\tmmathbf{p} \rangle  \label{eq:Gamsq}
\end{eqnarray}
with
\begin{eqnarray}
  \Gamma_0 \left( \tmmathbf{p} \right) & \assign &
  \frac{n_{\tmop{gas}}}{m_{\ast}} \left| \tmmathbf{p} \right|
  \sigma_{\tmop{tot}} \left( \tmmathbf{p} \right) .  \label{eq:Gamzero}
\end{eqnarray}
Noting that the $\mathsf{T}_0$ matrix elements are given by (\ref{eq:Tme}) one
thus obtains the generalized function
\begin{align}
  m_{\tmop{in}} \left( \tmmathbf{p}_f, \tmmathbf{p}_i ; \tmmathbf{q} \right) 
  &=  \Gamma_0^{1 / 2} \left( \tmmathbf{p}_i +\tmmathbf{q} \right) \Gamma_0^{1
  / 2} \left( \tmmathbf{p}_i -\tmmathbf{q} \right) \nonumber\\
    & \times f \left( \tmmathbf{p}_f +\tmmathbf{q}, \tmmathbf{p}_i
  +\tmmathbf{q} \right) f^{\ast} \left( \tmmathbf{p}_f -\tmmathbf{q},
  \tmmathbf{p}_i -\tmmathbf{q} \right) \nonumber\\
    & \times \frac{2 \pi \hbar}{| \Omega |} \delta \left(
  \frac{\tmmathbf{p}_f^2 -\tmmathbf{p}_i^2}{2} + \left( \tmmathbf{p}_f
  -\tmmathbf{p}_i \right) \cdot \tmmathbf{q} \right) \nonumber\\
    & \times \delta \left( \frac{\tmmathbf{p}_f^2 -\tmmathbf{p}_i^2}{2} -
  \left( \tmmathbf{p}_f -\tmmathbf{p}_i \right) \cdot \tmmathbf{q} \right) . 
  \label{eq:mrel1}
\end{align}
Clearly, the two delta-functions ensure that the energy is conserved in each
of the ``elastic collision trajectories'' expressed by the arguments of the
two scattering amplitudes in (\ref{eq:mrel1}). Employing the relation $\delta
\left( a + b / 2 \right) \delta \left( a - b / 2 \right) = \delta \left( a
\right) \delta \left( b \right)$ we obtain the equivalent form
\begin{align}
  m_{\tmop{in}} \left( \tmmathbf{p}_f, \tmmathbf{p}_i ; \tmmathbf{q} \right) 
  = & \delta \left( \frac{\tmmathbf{p}_f^2 -\tmmathbf{p}_i^2}{2} \right)
  \Gamma_0^{1 / 2} \left( \tmmathbf{p}_i +\tmmathbf{q} \right) \Gamma_0^{1 /
  2} \left( \tmmathbf{p}_i -\tmmathbf{q} \right) \nonumber\\
    & \times f \left( \tmmathbf{p}_f +\tmmathbf{q}, \tmmathbf{p}_i
  +\tmmathbf{q} \right) f^{\ast} \left( \tmmathbf{p}_f -\tmmathbf{q},
  \tmmathbf{p}_i -\tmmathbf{q} \right)  \nonumber\\
    & \times \delta \left(  \left( \tmmathbf{p}_f -\tmmathbf{p}_i \right)
  \cdot \tmmathbf{q} \right)  \frac{\pi \hbar}{| \Omega |} .  \label{eq:mrel2}
\end{align}
The first delta function now requires $\tmmathbf{p}_i$ and $\tmmathbf{p}_f$ to
have equal length. These are the mean relative momenta of the pairs of
scattering trajectories, as defined in Eqs.~(\ref{eq:pidef}) and
(\ref{eq:pfdef}). Given $\left| \tmmathbf{p}_i \right| = \left| \tmmathbf{p}_f
\right|$, the second delta function ensures that the energy is conserved in
each of the scattering amplitudes individually, by granting that
$\tmmathbf{q}$, which expresses a ``distance'' between the two pairs of
scattering trajectories, is orthogonal to the momentum exchange
$\tmmathbf{p}_i -\tmmathbf{p}_f$. The fact that possible parallel components
of $\tmmathbf{q}= {\tmmathbf{q}_{\bot} +\tmmathbf{q}_{\|}}$ cannot contribute
to an integral over the generalized function (\ref{eq:mrel2}) can be made
manifest by replacing the $\tmmathbf{q}$'s outside of the delta function by
the orthogonal component $\tmmathbf{q}_{\bot}$ defined in (\ref{eq:qbot}). In
other words, the statement (\ref{eq:mrel2}) is tantamount to
\begin{eqnarray}
  &  & m_{\tmop{in}} \left( \tmmathbf{p}_f, \tmmathbf{p}_i ; \tmmathbf{q}
  \right) \nonumber\\
  &  & = \hspace{0.6em} \delta \left( \frac{\tmmathbf{p}_f^2
  -\tmmathbf{p}_i^2}{2} \right) \Gamma_0^{1 / 2} \left( \tmmathbf{p}_i
  +\tmmathbf{q}_{\bot} \right) \Gamma_0^{1 / 2} \left( \tmmathbf{p}_i
  -\tmmathbf{q}_{\bot} \right) \nonumber\\
  &  & \phantom{= \hspace{0.6em}} \times f \left( \tmmathbf{p}_f
  +\tmmathbf{q}_{\bot}, \tmmathbf{p}_i +\tmmathbf{q}_{\bot} \right) f^{\ast}
  \left( \tmmathbf{p}_f -\tmmathbf{q}_{\bot}, \tmmathbf{p}_i
  -\tmmathbf{q}_{\bot} \right)  \nonumber\\
  &  &  \phantom{= \hspace{0.6em}} \times \delta \left(  \left(
  \tmmathbf{p}_f -\tmmathbf{p}_i \right) \cdot \tmmathbf{q} \right)  \frac{\pi
  \hbar}{| \Omega |},  \label{eq:mrel3}
\end{eqnarray}
which implies that the expression vanishes for $\tmmathbf{q}_{\|} \neq 0$, as
stated in (\ref{eq:mclaim0}). In fact, even if the integration over
$m_{\tmop{in}}$ involves a smooth function $g \left( \tmmathbf{q} \right)$ the
second delta function will enforce that the latter contributes only with the
orthogonal component of $\tmmathbf{q}$,
\begin{eqnarray}
  m_{\tmop{in}} \left( \tmmathbf{p}_f, \tmmathbf{p}_i ; \tmmathbf{q} \right) g
  \left( \tmmathbf{q} \right) & = & m_{\tmop{in}} \left( \tmmathbf{p}_f,
  \tmmathbf{p}_i ; \tmmathbf{q}_{\bot} \right) g \left( \tmmathbf{q}_{\bot}
  \right) .  \label{eq:mrelprop}
\end{eqnarray}
One observes on the right hand side of (\ref{eq:mrel3}) that already the first
two lines now manifestly ensure the energy conservation of the pair of
collision trajectories described by the two scattering amplitudes. This is a
crucial requirement since the scattering amplitudes are not defined off the
energy shell (notwithstanding the fact that analytic continuations are often
considered and helpful in scattering theory). At the same time this implies
that the physical relevance of the second delta-function has been accounted
for once the parallel components of $\tmmathbf{q}$ have been set to zero.
Hence, the third line in (\ref{eq:mrel3}) is essentially dispensable, which is
all the more important since it renders $m_{\tmop{in}}$ an ill-defined
expression due to the appearance of the arbitrarily large normalization volume
$| \Omega |$.

As is well understood, the evaluation carried out in this subsection does not
yield a well-behaved result because it takes the momentum-diagonal form
(\ref{eq:Gamrel}) of the rate operator $\mathsf{\Gamma}$ too seriously. It was
already discussed in Sect. \ref{sec:operators} that either $\mathsf{\Gamma}$
should involve a projection to the subspace of incoming wave packets, or that
the operator $\mathsf{S}$ should be redefined such that it keeps the outgoing
wave packets unchanged. These two strategies will be implemented in
Sects.~\ref{sec:wp} and \ref{sec:rr}, yielding identical results. As one
expects, the overall structure of (\ref{eq:mrel3}), which is dictated by the
energy conservation, will not change, but the third line will be replaced by a
proper normalization.

\section{Wave packet evaluation}\label{sec:wp}

The aim of this section is to evaluate the generalized function
\begin{eqnarray}
  m_{\tmop{in}} \left( \tmmathbf{p}_f, \tmmathbf{p}_i ; \tmmathbf{q}_{\bot}
  \right) & = & \frac{\left( 2 \pi \hbar \right)^3}{| \Omega |} \langle
  \tmmathbf{p}_f +\tmmathbf{q}_{\bot} | \mathsf{T}_0 \mathsf{\Gamma}^{1 / 2}_0
  |\tmmathbf{p}_i +\tmmathbf{q}_{\bot} \rangle \nonumber\\
  &  & \times \langle \tmmathbf{p}_i -\tmmathbf{q}_{\bot} |
  \mathsf{\Gamma}_0^{1 / 2} \mathsf{T^{\dag}_0} |\tmmathbf{p}_f
  -\tmmathbf{q}_{\bot} \rangle  \label{eq:mrel4}
\end{eqnarray}
by consistently incorporating the fact that the rate operator
$\mathsf{\Gamma}_0$ should have a vanishing expectation value for those states
of the relative motion that are not of incoming type. As a first step, we will
write (\ref{eq:mrel4}) as the expectation value of a non-hermitian operator
with respect to a properly normalized momentum state of the relative motion.
For that purpose it is convenient to introduce the operator $\mathsf{Z}_0
\assign \mathsf{T}_0 \mathsf{\Gamma}^{1 / 2}_0$ and its translation by the
momentum $\tmmathbf{q}_{\bot}$,
\begin{eqnarray}
  \mathsf{Z}_{\tmmathbf{q}_{\bot}} & = & \exp \left( - \mathi
  \frac{\mathsf{x}_{\tmop{rel}} \cdot \tmmathbf{q}_{\bot}}{\hbar} \right)
  \mathsf{T}_0 \mathsf{\Gamma}^{1 / 2}_0 \exp \left( \mathi
  \frac{\mathsf{x}_{\tmop{rel}} \cdot \tmmathbf{q}_{\bot}}{\hbar} \right), 
  \label{eq:Zdef}
\end{eqnarray}
where $\mathsf{x}_{\tmop{rel}}$ is the position operator of the relative
coordinate. Moreover, we note that, analogous to (\ref{eq:rhopure}), a
volume-normalized momentum state of the{\tmem{ relative}} motion has the form
\begin{eqnarray}
  \rho_{\tmmathbf{p}_i} & = & \frac{\left( 2 \pi \hbar \right)^3}{| \Omega |}
  \mathsf{I}_{\Omega} |\tmmathbf{p}_i \rangle \langle \tmmathbf{p}_i |
  \mathsf{I}_{\Omega} .  \label{eq:rhorelpure}
\end{eqnarray}
Combining (\ref{eq:Zdef}) and (\ref{eq:rhorelpure}) one finds that the complex
rate density (\ref{eq:mrel4}) can be taken as the diagonal momentum matrix
element of an operator product, ${m_{\tmop{in}} \left( \tmmathbf{p}_f,
\tmmathbf{p}_i ; \tmmathbf{q}_{\bot} \right)} = {\langle \tmmathbf{p}_f |
\mathsf{Z}_{\tmmathbf{q}_{\bot}} \rho_{\tmmathbf{p}_i}
\mathsf{Z}_{-\tmmathbf{q}_{\bot}}^{\dag} |\tmmathbf{p}_f \rangle}$, provided
the projection to the normalization volume is included. If we further denote
the projector to improper momentum eigenstates as $\mathsf{P}_{\tmmathbf{p}_f}
= |\tmmathbf{p}_f \rangle \langle \tmmathbf{p}_f |$ we can write
\begin{eqnarray}
  m_{\tmop{in}} \left( \tmmathbf{p}_f, \tmmathbf{p}_i ; \tmmathbf{q}_{\bot}
  \right) & = & \tmop{Tr} \left( \mathsf{Z}_{-\tmmathbf{q}_{\bot}}^{\dag}
  \mathsf{P}_{\tmmathbf{p}_f}  \mathsf{Z}_{\tmmathbf{q}_{\bot}}
  \rho_{\tmmathbf{p}_i} \right) .  \label{eq:mrel6}
\end{eqnarray}
The complex rate density (\ref{eq:mrel4}) has now the form of an expectation
value with respect to a state $\rho_{\tmmathbf{p}_i}$ which is properly
normalized, $\tmop{Tr} \left( \rho_{\tmmathbf{p}_i} \right) = 1$. As discussed
in Sect. \ref{sec:operators}, the rate operator $\mathsf{\Gamma}_0$, and
therefore also the non-hermitian operator
$\mathsf{Z}_{-\tmmathbf{q}_{\bot}}^{\dag} \mathsf{P}_{\tmmathbf{p}_f} 
\mathsf{Z}_{\tmmathbf{q}_{\bot}}$ should include a restriction to the subspace
of truly incoming relative motional states. Starting from (\ref{eq:mrel6}),
one can now implement this restriction in a rather transparent and intuitive
fashion by considering the phase space representation of
$\rho_{\tmmathbf{p}_i}$, as shown in the next subsection. A similar method was
already successfully applied in {\cite{Hornberger2007b}}, where the effect of
a gas on the internal dynamics of an immobile system was discussed by
combining the monitoring approach with scattering theory.

\subsection{Phase space restriction to incoming wave
packets}\begin{figure}[tb]
  \resizebox{85mm}{!}{\epsfig{file=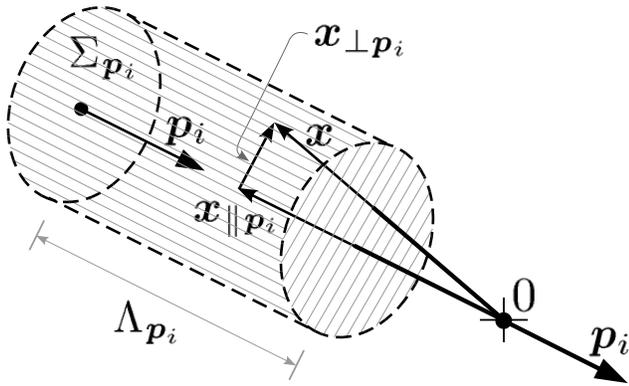}}
  \caption{\label{fig:cylinder}The incoming relative momentum $\tmmathbf{p}_i$
  defines a cylinder with base area $\Sigma_{\tmmathbf{p}_i}$ and height
  $\Lambda_{\tmmathbf{p}_i}$. This spatial region is used to implement the
  phase space restriction of the Wigner function to incoming states. }
\end{figure}

The operator $\rho_{\tmmathbf{p}_i}$ in (\ref{eq:mrel6}) characterizes the
motional state of the relative coordinates between particle and gas prior to a
collision. According to (\ref{eq:rhorelpure}) it is given by a plane wave
which extends through the whole normalization volume, and it is therefore
clearly not of the incoming type required for the application of scattering
theory.

The Wigner-Weyl formulation of quantum mechanics
{\cite{Hillery1984a,Ozorio1998a,Schleich2001a,Zachos2005a}} suggests a way to
treat this problem. Continuous variable states may be represented by the phase
space quasi-probability function $W_{\rho} \left( \tmmathbf{x}, \tmmathbf{p}
\right) \assign \left( 2 \pi \hbar \right)^{- 3} \int \hspace{-0.25em} \mathd
\tmmathbf{q} \exp \left( \um \mathi \hspace{0.25em} \tmmathbf{q} \cdot
\tmmathbf{x}/ \hbar \right) \hspace{0.25em} \langle \tmmathbf{p}-\tmmathbf{q}/
2| \rho |\tmmathbf{p}+\tmmathbf{q}/ 2 \rangle$. For the state
(\ref{eq:rhorelpure}) the associated Wigner function reads{\footnote{This
follows with the approximation $\chi_{\Omega} \left( \tmmathbf{r}-
\frac{\tmmathbf{s}}{2} \right) \chi_{\Omega} \left( \tmmathbf{r}+
\frac{\tmmathbf{s}}{2} \right) \simeq \chi_{\Omega} \left( \tmmathbf{r}
\right) \chi_{\Omega} \left( \tmmathbf{s} \right)$, which is permissible since
the normalization region $\Omega$ may be taken arbitrarily large.}}
\begin{eqnarray}
  W_{\tmmathbf{p}_i} \left( \tmmathbf{x}, \tmmathbf{p} \right) & = &
  \frac{\chi_{\Omega} \left( \tmmathbf{x} \right)}{| \Omega |} \delta \left(
  \tmmathbf{p}-\tmmathbf{p}_i \right),  \label{eq:Wigner1}
\end{eqnarray}
where $\chi_{\Omega}$ is the characteristic function of the normalization
volume $| \Omega |$. Given that expectation values like (\ref{eq:mrel6}) may
now be calculated as phase space integrals, it is natural to implement the
restriction by confining (\ref{eq:Wigner1}) to the phase space area of
incoming wave packets, i.e.,
\begin{eqnarray}
  W'_{\tmmathbf{p}_i} \left( \tmmathbf{x}, \tmmathbf{p} \right) & = &
  \frac{\chi_{\Lambda_{\tmmathbf{p}_i}} \left( \tmmathbf{x}_{\|\tmmathbf{p}_i}
  \right)}{| \Lambda_{\tmmathbf{p}_i} |} \frac{\chi_{\Sigma_{\tmmathbf{p}_i}}
  \left( \tmmathbf{x}_{\bot \tmmathbf{p}_i} \right)}{| \Sigma_{\tmmathbf{p}_i}
  |} \delta \left( \tmmathbf{p}-\tmmathbf{p}_i \right) .  \label{eq:Wigner2}
\end{eqnarray}
Here, the product $\chi_{\Lambda_{\tmmathbf{p}_i}} \left(
\tmmathbf{x}_{\|\tmmathbf{p}_i} \right) \chi_{\Sigma_{\tmmathbf{p}_i}} \left(
\tmmathbf{x}_{\bot \tmmathbf{p}_i} \right) $is the characteristic function of
a cylinder pointing towards the origin, see Fig.~\ref{fig:cylinder}. It
describes those points $\tmmathbf{x}=\tmmathbf{x}_{\|\tmmathbf{p}_i}
+\tmmathbf{x}_{\bot \tmmathbf{p}_i}$ in position space which will pass the
vicinity of the origin when propagated in the direction given by
$\tmmathbf{p}_i$. $\Sigma_{\tmmathbf{p}_i}$ is the base surface of the
cylinder and its area will be taken to be equal to the total cross section
below, i.e., $| \Sigma_{\tmmathbf{p}_i} | = \sigma \left( \tmmathbf{p}_i
\right)$. The interval $\Lambda_{\tmmathbf{p}_i} \mathbbm{}$ specifies the
cylinder height; its precise $\tmmathbf{p}_i$-dependence will drop out of the
calculation later on.

The operator corresponding to (\ref{eq:Wigner2}) reads
\begin{eqnarray}
  \rho'_{\tmmathbf{p}_i} & = & \int_{\Lambda_{\tmmathbf{p}_i}} \frac{\mathd
  \tmmathbf{x}_{\|\tmmathbf{p}_i}}{| \Lambda_{\tmmathbf{p}_i} |}
  \int_{\Sigma_{\tmmathbf{p}_i}} \frac{\mathd \tmmathbf{x}_{\bot
  \tmmathbf{p}_i}}{| \Sigma_{\tmmathbf{p}_i} |} \bignone \int \mathd
  \tmmathbf{w} \, \nonumber\\
  &  & \times \exp \left( i \frac{\tmmathbf{x} \cdot \tmmathbf{w}}{\hbar}
  \right) |\tmmathbf{p}_i - \frac{\tmmathbf{w}}{2} \rangle \langle
  \tmmathbf{p}_i + \frac{\tmmathbf{w}}{2} |,  \label{eq:rhopip}
\end{eqnarray}
so that compared to the unrestricted expression corresponding to
(\ref{eq:Wigner1})
\begin{eqnarray}
  &  & \int_{\Omega} \frac{\mathd \tmmathbf{x}_{_{}}}{| \Omega |} \bignone
  \int \mathd \tmmathbf{w} \, \exp \left( i \frac{\tmmathbf{x} \cdot
  \tmmathbf{w}}{\hbar} \right) |\tmmathbf{p}_i - \frac{\tmmathbf{w}}{2}
  \rangle \langle \tmmathbf{p}_i + \frac{\tmmathbf{w}}{2} |, 
\end{eqnarray}
the spatial average over the whole normalization volume is simply replaced by
an average over the cylinder, and the norm is indeed preserved,
\begin{eqnarray}
  \tmop{Tr} \left( \rho'_{\tmmathbf{p}_i} \right) & = &
  \int_{\Lambda_{\tmmathbf{p}_i}} \frac{\mathd
  \tmmathbf{x}_{\|\tmmathbf{p}_i}}{| \Lambda_{\tmmathbf{p}_i} |}
  \int_{\Sigma_{\tmmathbf{p}_i}} \frac{\mathd \tmmathbf{x}_{\bot
  \tmmathbf{p}_i}}{| \Sigma_{\tmmathbf{p}_i} |} \bignone \hspace{0.6em} =
  \hspace{0.6em} 1. 
\end{eqnarray}
It should be noted, though, that strictly speaking neither
$\rho_{\tmmathbf{p}_i}$ nor $\rho_{\tmmathbf{p}_i}'$ are legitimate quantum
states since they combine a precise momentum with a finite position variance.
They should rather be seen as convenient basis states admitting to average
over the momentum distribution function, see Eqs.~(\ref{eq:rhogasmu}) and
(\ref{eq:Minrel}).

\subsection{Restricted evaluation}\label{sec:resev}

Inserting the restricted state (\ref{eq:rhopip}) into (\ref{eq:mrel6}) one can
now evaluate the complex rate density to obtain a well-defined expression.
Starting with
\begin{eqnarray}
  &  & m_{\tmop{in}} \left( \tmmathbf{p}_f, \tmmathbf{p}_i ;
  \tmmathbf{q}_{\bot} \right) \nonumber\\
  &  & \cong \hspace{0.6em} \tmop{Tr} \left(
  \mathsf{Z}_{-\tmmathbf{q}_{\bot}}^{\dag} \mathsf{P}_{\tmmathbf{p}_f} 
  \mathsf{Z}_{\tmmathbf{q}_{\bot}} \rho_{\tmmathbf{p}_i}' \right) \nonumber\\
  &  & = \hspace{0.6em} \int_{\Lambda_{\tmmathbf{p}_i}} \frac{\mathd
  \tmmathbf{x}_{\|\tmmathbf{p}_i}}{| \Lambda_{\tmmathbf{p}_i} |}
  \int_{\Sigma_{\tmmathbf{p}_i}} \frac{\mathd \tmmathbf{x}_{\bot
  \tmmathbf{p}_i}}{| \Sigma_{\tmmathbf{p}_i} |} \bignone \int \mathd
  \tmmathbf{w} \, \exp \left( i \frac{\tmmathbf{x} \cdot \tmmathbf{w}}{\hbar}
  \right) \nonumber\\
  &  & \phantom{= \hspace{0.6em}} \times \langle \tmmathbf{p}_f
  +\tmmathbf{q}_{\bot} | \mathsf{T}_0 \mathsf{\Gamma}^{1 / 2}_0
  |\tmmathbf{p}_i +\tmmathbf{q}_{\bot} - \frac{\tmmathbf{w}}{2} \rangle
  \nonumber\\
  &  & \phantom{= \hspace{0.6em}} \times \langle \tmmathbf{p}_i
  -\tmmathbf{q}_{\bot} + \frac{\tmmathbf{w}}{2} | \mathsf{\Gamma}_0^{1 / 2}
  \mathsf{T^{\dag}_0} |\tmmathbf{p}_f -\tmmathbf{q}_{\bot} \rangle 
  \label{eq:mrel7}
\end{eqnarray}
we can now use with confidence the expressions (\ref{eq:Tme}) and
(\ref{eq:Gamsq}) for the momentum matrix elements of $\mathsf{T}_0 \tmop{and}
\mathsf{\Gamma}^{1 / 2}_0$. Thus, $m_{\tmop{in}} \left( \tmmathbf{p}_f,
\tmmathbf{p}_i ; \tmmathbf{q}_{\bot} \right)$ takes the form
\begin{eqnarray}
  &  & m_{\tmop{in}} \left( \tmmathbf{p}_f, \tmmathbf{p}_i ;
  \tmmathbf{q}_{\bot} \right) \nonumber\\
  &  & = \hspace{0.6em}  \int_{\Lambda_{\tmmathbf{p}_i}} \frac{\mathd
  \tmmathbf{x}_{\|\tmmathbf{p}_i}}{| \Lambda_{\tmmathbf{p}_i} |}
  \int_{\Sigma_{\tmmathbf{p}_i}} \frac{\mathd \tmmathbf{x}_{\bot
  \tmmathbf{p}_i}}{| \Sigma_{\tmmathbf{p}_i} |} \bignone \int \mathd
  \tmmathbf{w} \, \exp \left( - i \frac{\tmmathbf{x} \cdot
  \tmmathbf{w}}{\hbar} \right) \nonumber\\
  &  & \phantom{= \hspace{0.6em}} \times \frac{1}{\left( 2 \pi \hbar
  \right)^2} f \left( \tmmathbf{p}_f +\tmmathbf{q}_{\bot}, \tmmathbf{p}_i
  +\tmmathbf{q}_{\bot} + \frac{\tmmathbf{w}}{2} \right) \nonumber\\
  &  & \phantom{= \hspace{0.6em}} \times f^{\ast} \left( \tmmathbf{p}_f
  -\tmmathbf{q}_{\bot}, \tmmathbf{p}_i -\tmmathbf{q}_{\bot} -
  \frac{\tmmathbf{w}}{2} \right) \nonumber\\
  &  & \phantom{= \hspace{0.6em}} \times \delta \left( \frac{\tmmathbf{p}_f^2
  -\tmmathbf{p}_i^2}{2} - \frac{\tmmathbf{q}_{\bot} \cdot \tmmathbf{w}}{2} -
  \frac{w^2}{8} - \frac{1}{2} \tmmathbf{p}_i \cdot \tmmathbf{w} \right)
  \nonumber\\
  &  & \phantom{= \hspace{0.6em}} \times \delta \left( \frac{\tmmathbf{p}_f^2
  -\tmmathbf{p}_i^2}{2} - \frac{\tmmathbf{q}_{\bot} \cdot \tmmathbf{w}}{2} -
  \frac{w^2}{8} + \frac{1}{2} \tmmathbf{p}_i \cdot \tmmathbf{w} \right)
  \nonumber\\
  &  & \phantom{= \hspace{0.6em}} \times \Gamma^{1 / 2}_0 \left(
  \tmmathbf{p}_i +\tmmathbf{q}_{\bot} + \frac{\tmmathbf{w}}{2} \right)
  \Gamma^{1 / 2}_0 \left( \tmmathbf{p}_i -\tmmathbf{q}_{\bot} -
  \frac{\tmmathbf{w}}{2} \right) . \nonumber\\
\end{eqnarray}
In the arguments of the delta function we took into account that
$\tmmathbf{q}_{\bot}$ is orthogonal to the momentum exchange, $\left(
\tmmathbf{p}_f -\tmmathbf{p}_i \right) \cdot \tmmathbf{q}_{\bot} = 0$, as
follows from (\ref{eq:qbot}). Using again the relation $\delta \left( a + b /
2 \right) \delta \left( a - b / 2 \right) = \delta \left( a \right) \delta
\left( b \right)$ a delta function is obtained with argument {$\tmmathbf{p}_i
\cdot \tmmathbf{w}$}. Writing {$\tmmathbf{w}=\tmmathbf{w}_{\|\tmmathbf{p}_i}
+\tmmathbf{w}_{\bot \tmmathbf{p}_i}$}, with $\tmmathbf{w}_{\|\tmmathbf{p}_i} =
\left( \tmmathbf{w} \cdot \tmmathbf{p}_i \right) \tmmathbf{p}_i / p^2_i$, this
delta function renders $\tmmathbf{w}_{\|\tmmathbf{p}_i} = 0$, and as a result
the integrand now no longer depends on $\tmmathbf{x}_{\|\tmmathbf{p}_i}$. It
follows that the integration along the cylinder axis can be
done,$\int_{\Lambda_{\tmmathbf{p}_i}} \mathd \tmmathbf{x}_{\|\tmmathbf{p}_i} =
| \Lambda_{\tmmathbf{p}_i} |$. We obtain
\begin{eqnarray}
  &  & m_{\tmop{in}} \left( \tmmathbf{p}_f, \tmmathbf{p}_i ;
  \tmmathbf{q}_{\bot} \right) \nonumber\\
  &  & = \hspace{0.6em}  \frac{n_{\tmop{gas}}}{m_{\ast}}  \int \mathd
  \tmmathbf{w} \, \delta \left(  \frac{\tmmathbf{p}_i \cdot \tmmathbf{w}}{p_i}
  \right) \bignone \int_{\Sigma_{\tmmathbf{p}_i}} \frac{\mathd
  \tmmathbf{x}_{\bot \tmmathbf{p}_i}}{\left( 2 \pi \hbar \right)^2} 
  \nonumber\\
  &  & \phantom{= \hspace{0.6em}} \times \exp \left( - i
  \frac{\tmmathbf{x}_{\bot \tmmathbf{p}_i} \cdot \tmmathbf{w}_{\bot
  \tmmathbf{p}_i}}{\hbar} \right) \bignone \nonumber\\
  &  & \phantom{= \hspace{0.6em}} \times \delta \left( \frac{\tmmathbf{p}_f^2
  -\tmmathbf{p}_i^2}{2} - \frac{\tmmathbf{q}_{\bot} \cdot \tmmathbf{w}}{2} -
  \frac{w^2}{8} \right) \nonumber\\
  &  & \phantom{= \hspace{0.6em}} \times f \left( \tmmathbf{p}_f
  +\tmmathbf{q}_{\bot}, \tmmathbf{p}_i +\tmmathbf{q}_{\bot} +
  \frac{\tmmathbf{w}}{2} \right) \nonumber\\
  &  & \phantom{= \hspace{0.6em}} \times f^{\ast} \left( \tmmathbf{p}_f
  -\tmmathbf{q}_{\bot}, \tmmathbf{p}_i -\tmmathbf{q}_{\bot} -
  \frac{\tmmathbf{w}}{2} \right) \nonumber\\
  &  & \phantom{= \hspace{0.6em}} \times \frac{1}{p_i} \sqrt{\left|
  \tmmathbf{p}_i +\tmmathbf{q}_{\bot} + \frac{\tmmathbf{w}}{2} \right| \left|
  \tmmathbf{p}_i -\tmmathbf{q}_{\bot} - \frac{\tmmathbf{w}}{2} \right|}
  \nonumber\\
  &  & \phantom{= \hspace{0.6em}} \times \frac{1}{\sigma_{\tmop{tot}} \left(
  \tmmathbf{p}_i \right)} \sqrt{\sigma_{\tmop{tot}} \left( \tmmathbf{p}_i
  +\tmmathbf{q}_{\bot} + \frac{\tmmathbf{w}}{2} \right)} \nonumber\\
  &  & \phantom{= \hspace{0.6em}} \times \sqrt{\sigma_{\tmop{tot}} \left(
  \tmmathbf{p}_i -\tmmathbf{q}_{\bot} - \frac{\tmmathbf{w}}{2} \right)}, 
\end{eqnarray}
where we identified the cylinder base area with the total cross section, $|
\Sigma_{\tmmathbf{p}_i} | = \sigma_{\tmop{tot}} \left( \tmmathbf{p}_i
\right)$. One observes that the $\tmmathbf{x}_{\bot
\tmmathbf{p}_i}$-integration over the surface $\Sigma_{\tmmathbf{p}_i}$ of the
cylinder base yields an approximate two-dimensional delta function in
$\tmmathbf{w}_{\bot \tmmathbf{p}_i}$. Combined with the one-dimensional delta
function in $w_{\|\tmmathbf{p}_i} =\tmmathbf{p}_i \cdot \tmmathbf{w}/ p_i$
this gives a three-dimensional $\delta \left( \tmmathbf{w} \right)$, which
permits to carry out the $\tmmathbf{w}$-integration. We arrive at the
well-defined expression
\begin{eqnarray}
  &  & m_{\tmop{in}} \left( \tmmathbf{p}_f, \tmmathbf{p}_i ;
  \tmmathbf{q}_{\bot} \right) \nonumber\\
  &  & = \hspace{0.6em}  \frac{n_{\tmop{gas}}}{m_{\ast}} \delta \left(
  \frac{\tmmathbf{p}_f^2 -\tmmathbf{p}_i^2}{2} \right) f \left( \tmmathbf{p}_f
  +\tmmathbf{q}_{\bot}, \tmmathbf{p}_i +\tmmathbf{q}_{\bot} \right)
  \nonumber\\
  &  & \phantom{= \hspace{0.6em}} \times f^{\ast} \left( \tmmathbf{p}_f
  -\tmmathbf{q}_{\bot}, \tmmathbf{p}_i -\tmmathbf{q}_{\bot} \right)
  \nonumber\\
  &  & \phantom{= \hspace{0.6em}} \times \frac{\sqrt{\left| \tmmathbf{p}_i
  +\tmmathbf{q}_{\bot} \right| \left| \tmmathbf{p}_i -\tmmathbf{q}_{\bot}
  \right|}}{p_i} \nonumber\\
  &  & \phantom{= \hspace{0.6em}} \times \frac{\sqrt{\sigma_{\tmop{tot}}
  \left( \tmmathbf{p}_i +\tmmathbf{q}_{\bot} \right) \sigma_{\tmop{tot}}
  \left( \tmmathbf{p}_i -\tmmathbf{q}_{\bot} \right)}}{\sigma_{\tmop{tot}}
  \left( \tmmathbf{p}_i \right)} .  \label{eq:mrel10}
\end{eqnarray}
It shows that the complex rate density (\ref{eq:Minrel}) is essentially given
by the product of two scattering amplitudes whose arguments differ in general.
They correspond to scattering ``trajectories'' determined by the relative
momenta $\tmmathbf{p}_i$, $\tmmathbf{p} _f$, and $\tmmathbf{q}_{\bot}$. The
$\tmmathbf{p}_i$ and $\tmmathbf{p} _f$ provide the arithmetic means of the
initial and the final momenta, while $\tmmathbf{q}_{\bot}$ characterizes the
distance of the ``trajectories'' in momentum space. Since
$\tmmathbf{q}_{\bot}$ is orthogonal to $\tmmathbf{p}_f -\tmmathbf{p}_i$, a
single delta-function suffices in (\ref{eq:mrel10}) to ensure the conservation
of energy in both scattering amplitudes.

Reassuringly, this result reduces to the classical rate density
(\ref{eq:Mclin}) for $\tmmathbf{q}_{\bot} = 0$, as can be seen easily by
inserting $m_{\tmop{in}} \left( \tmmathbf{p}_f, \tmmathbf{p}_i ; 0 \right)$
into (\ref{eq:Minrel}). This shows that the first line in (\ref{eq:mrel10})
may be viewed as a natural quantum generalization of the classical case, where
the ``off-diagonal'' contributions with $\tmmathbf{q}_{\bot} \neq 0$ represent
quantum corrections. From the point of view of quantum physics, there is
indeed no reason why the effect of the gas collisions on the tracer particle
should be confined to the ``diagonal'' contributions given by
$\tmmathbf{q}_{\bot} = 0$. In the present relative coordinate representation,
the first line in (\ref{eq:mrel10}) has in fact a straightforward
interpretation. It simply provides the contribution of the scattering
amplitudes of any pair of scattering trajectories, which is allowed by both
the energy conservation and the choice of $\tmmathbf{P}, \tmmathbf{P}'$, and
$\tmmathbf{Q}$ in (\ref{eq:Mexp}).

At the same time, one expects that a $\tmmathbf{q}_{\bot}$-integration will
average out the ``far off-diagonal'' contributions with large modulus $\left|
\tmmathbf{q}_{\bot} \right|$, where the phases of the two scattering
amplitudes are no longer synchronous. It is therefore reasonable to disregard
the weak $\tmmathbf{q}_{\bot}$-dependence in the second line of
(\ref{eq:mrel10}), and this is corroborated by the fact that its linear
dependence on $\tmmathbf{q}_{\bot}$ vanishes identically. This removes the
second line altogether, so that we end up with the form claimed in
Eq.~(\ref{eq:mclaim}).

A noteworthy step in the present line of reasoning was that the base area $|
\Sigma_{\tmmathbf{p}_i} |$ of the cylinder required for distinguishing the
incoming states was identified with the scattering cross section
$\sigma_{\tmop{tot}}$. This is very natural from a physical point of view, but
it seems hard to justify on a formal basis. It is therefore worthwhile to
present a second argumentation which, though very different in nature, leads
to the identical result.

\section{Evaluation with modified scattering
operator}\label{sec:rr}

A second, more heuristic approach of evaluating $m_{\tmop{in}}$ sidesteps the
issue of how to incorporate a restriction of the rate operator
$\mathsf{\Gamma}_0$ to the incoming wave packets, and takes its
momentum-diagonal form (\ref{eq:Gamrel}) at face value. The unrestricted
$\mathsf{\Gamma}_0$ then attributes a finite scattering rate also to{\tmem{
outgoing}} wave packets, which would never touch the interaction region in a
dynamic description. This forces us to consider a redefinition of the
scattering operator, which is necessary since outgoing wave packets are not
left invariant by the proper S-matrix $\mathsf{S}_0$, as discussed above in
Fig.~\ref{fig:cartoon}. Let us therefore formally replace $\mathsf{S}_0$ by a
modified operator ${\mathsf{S}_0'}$, which by construction acts like
$\mathsf{S}_0$ when applied to asymptotic-in states, but leaves states with
outgoing characteristics invariant. It will not be necessary to specify the
details of this modification since all that is needed for the evaluation of
$m_{\tmop{in}}$ can be obtained from a single property that must hold for any
such modified operator. It is the isometry of $\mathsf{S}_0'$ with respect to
the set of volume-normalized momentum states.

The advantage of replacing ${\mathsf{S}_0}$ by ${\mathsf{S}_0'}$ is that we
can now use plane waves not merely as a basis, but as representing proper
states, because the unwanted transformation of their outgoing components is
now formally excluded. Like in Sect.~\ref{sec:monitoring}, we use double
brackets to denote momentum states which are normalized with respect to the
volume $\Omega$ and subject to periodic boundary conditions on its border. Due
to the finite size of $\Omega$ they form a discrete basis $\left\{
||\tmmathbf{p} \rangle \rangle : \tmmathbf{p} \in \mathbbm{P_{\Omega}}
\right\}$, decomposing the identity (\ref{eq:Omegaprojection}) as
\begin{eqnarray}
  \mathsf{I}_{\Omega} & = & \sum_{\tmmathbf{p} \in \mathbbm{P}_{\Omega}}
  ||\tmmathbf{p} \rangle \rangle \langle \langle \tmmathbf{p}||. 
  \label{eq:Idisc}
\end{eqnarray}
Heuristically, one may view each $\tmmathbf{p} \in \mathbbm{P_{\Omega}}$ as
labeling a distinct lead connected into and out of the scattering center. An
important difference with respect to a continuous description is that the
unitarity of the proper S-matrix, which expresses itself in the optical
theorem, cannot be accommodated within this discrete setting. The optical
theorem quantifies the diffraction limitation of the scattering cross section,
telling `how much' of a plane wave would pass the scattering center without
distortion. If we describe the scattering process in terms of the amplitudes
corresponding to discrete momentum states, or distinct leads, then the
possibility of `passing the target' is no longer available since any matrix
element may have a finite amplitude. [The possibility of `forward scattering'
$||\tmmathbf{p}_i \rangle \rangle \rightarrow ||\tmmathbf{p}_i \rangle
\rangle$ differs from this diffractive ``passing'' and leads to an additional
phase shift, see the following section.] This suggests to disregard the
identity operator in $\mathsf{S}_0'$, which relates to the unscattered part of
the state, and to require of the remaining transition operator that it
conserves the norm,
\begin{eqnarray}
  \| \mathsf{S}_0' ||\tmmathbf{p}_i \rangle \rangle \|^2 \hspace{0.6em} =
  \hspace{0.6em} \| \mathsf{T}_0' ||\tmmathbf{p}_i \rangle \rangle \|^2  & = &
  1. 
\end{eqnarray}
Inserting the identity (\ref{eq:Idisc}) we see that the sum of the
probabilities of scattering into the different leads equals 1,
\begin{eqnarray}
  \sum_{\tmmathbf{p} \in \mathbbm{P}_{\Omega}} | \langle \langle \tmmathbf{p}|
  \mathsf{T}'_0 |\tmmathbf{p}_i \rangle \rangle |^2 & = & 1. 
  \label{eq:discone}
\end{eqnarray}
This is the standard property of the transition matrix used to describe
discrete scattering problems between a finite number of incoming and outgoing
leads, e.g. in mesoscopic physics {\cite{Imry1997a}} or the field of quantum
graphs {\cite{Kottos2003a}}. It seems natural to demand this relation of any
reasonably modified operator$\mathsf{T}_0'$.

The use of (\ref{eq:discone}) is that it tells us how to normalize the square
of $\mathsf{T}_0'$ matrix elements with respect to improper momentum kets.
Inspecting the momentum matrix element of the $\mathsf{T}_0$ operator given in
(\ref{eq:Tme}) one finds that $| \langle \tmmathbf{p}_f | \mathsf{T}_0
|\tmmathbf{p}_i \rangle |^2$ involves the square of a delta-function. The
expression should be well-defined when using the modified operator
$\mathsf{T}_0'$, and the obvious choice is to assume that it is given by the
corresponding expression with a single delta-function and a normalization
$\mathcal{N} \left( \tmmathbf{p}_i \right)$ yet to be specified,
\begin{eqnarray}
  \frac{\left( 2 \pi \hbar \right)^3}{| \Omega |} | \langle \tmmathbf{p}_f |
  \mathsf{T}_0' |\tmmathbf{p}_i \rangle |^2 \bignone & = & \mathcal{N} \left(
  \tmmathbf{p}_i \right) \delta \left( \frac{\tmmathbf{p}^2_f
  -\tmmathbf{p}^2_i}{2} \right) \left| f \left( \tmmathbf{p}_f, \tmmathbf{p}_i
  \right) \right|^2 . \nonumber\\
  &  & 
\end{eqnarray}
Approximating the summation in (\ref{eq:discone}) by the corresponding
integral one finds
\begin{eqnarray}
  1 & = & \int \mathd \tmmathbf{p}_f  \frac{\left( 2 \pi \hbar \right)^3}{|
  \Omega |} | \langle \tmmathbf{p}_f | \mathsf{T}_0' |\tmmathbf{p}_i \rangle
  |^2 \bignone \nonumber\\
  & = & \mathcal{N} \left( \tmmathbf{p}_i \right)  \int \mathd \tmmathbf{p}_f
  \, \delta \left( \frac{\tmmathbf{p}^2_f -\tmmathbf{p}^2_i}{2} \right)
  \bignone |f \left( \tmmathbf{p}_f, \tmmathbf{p}_i \right) |^2 \nonumber\\
  & = & \mathcal{N} \left( \tmmathbf{p}_i \right) |\tmmathbf{p}_i |
  \sigma_{\tmop{tot}} \left( \tmmathbf{p}_i \right) . \nonumber
\end{eqnarray}
This fixes the normalization, $\mathcal{N} \left( \tmmathbf{p}_i \right) =
\left[ |\tmmathbf{p}_i | \sigma_{\tmop{tot}} \left( \tmmathbf{p}_i \right)
\right]^{- 1}$ and we obtain a well-defined expression for the squared matrix
element of the modified operator $\mathsf{T}_0'$,
\begin{eqnarray}
  \frac{\left( 2 \pi \hbar \right)^3}{| \Omega |} | \langle \tmmathbf{p}_f |
  \mathsf{T}_0' |\tmmathbf{p}_i \rangle |^2 & = & \delta \left(
  \frac{\tmmathbf{p}_f^2 -\tmmathbf{p}^2_i}{2} \right) \frac{\left| f \left(
  \tmmathbf{p}_f, \tmmathbf{p}_i \right) \right|^2}{\sigma_{\tmop{tot}} \left(
  \tmmathbf{p}_i \right)  \left| \tmmathbf{p}_i \right|} . \nonumber\\
  &  &  \label{eq:replacementrule}
\end{eqnarray}
Arriving at this equation required a certain amount of heuristic
argumentation. It should be emphasized that this expression was already used
in {\cite{Hornberger2003b}}, where it was shown to yield a localization rate
for collisional decoherence that is equal to a wave packet calculation similar
to the one in Sect.~\ref{sec:wp}.

If we accept (\ref{eq:replacementrule}) the evaluation of the complex rate
density $m_{\tmop{in}}$ can be done in a rather straightforward fashion. Using
the unrestricted rate operator $\mathsf{\Gamma}_0$ and the modified
$\mathsf{T}_0'$ instead of $\mathsf{T}_0$, the complex rate density from
(\ref{eq:mreldef}) takes the form
\begin{eqnarray}
  m_{\tmop{in}} \left( \tmmathbf{p}_f, \tmmathbf{p}_i ; \tmmathbf{q} \right) &
  = & \Gamma_0^{1 / 2} \left( \tmmathbf{p}_i +\tmmathbf{q} \right) \Gamma_0^{1
  / 2} \left( \tmmathbf{p}_i -\tmmathbf{q} \right) \xi \left( \tmmathbf{p}_f,
  \tmmathbf{p}_i ; \tmmathbf{q} \right) \nonumber\\
  &  &  \label{eq:minxi}
\end{eqnarray}
with the formal expression
\begin{eqnarray}
  \xi \left( \tmmathbf{p}_f, \tmmathbf{p}_i ; \tmmathbf{q} \right) & = &
  \frac{\left( 2 \pi \hbar \right)^3}{| \Omega |} \langle \tmmathbf{p}_f
  +\tmmathbf{q}| \mathsf{T}_0' |\tmmathbf{p}_i +\tmmathbf{q} \rangle
  \nonumber\\
  &  & \times \langle \tmmathbf{p}_f -\tmmathbf{q}| \mathsf{T}'_0
  |\tmmathbf{p}_i -\tmmathbf{q} \rangle^{\ast} .  \label{eq:xidef}
\end{eqnarray}
The latter can be evaluated by means of Eq.~(\ref{eq:replacementrule}). For
$\tmmathbf{q}= 0$ we have immediately
\begin{eqnarray}
  \xi \left( \tmmathbf{p}_f, \tmmathbf{p}_i ; 0 \right) & = & \delta \left(
  \frac{\tmmathbf{p}_f^2 -\tmmathbf{p}^2_i}{2} \right) \frac{\left| f \left(
  \tmmathbf{p}_f, \tmmathbf{p}_i \right) \right|^2}{\sigma_{\tmop{tot}} \left(
  \tmmathbf{p}_i \right)  \left| \tmmathbf{p}_i \right|}, 
\end{eqnarray}
while for $\tmmathbf{q} \neq 0$ an extension of the rule
(\ref{eq:replacementrule}) to different pairs of incoming and outgoing
relative momenta is required. It can be constructed by formally taking the
square root of (\ref{eq:replacementrule}). Insertion into (\ref{eq:xidef})
brings about the square root of a product of two energy conserving
$\delta$-functions with arguments $ \frac{\tmmathbf{p}_f^2
-\tmmathbf{p}_i^2}{2} \pm \left( \tmmathbf{p}_f -\tmmathbf{p}_i \right) \cdot
\tmmathbf{q}$. Like with the delta-functions in Sect.~\ref{sec:resev}, this
product implies that the parallel component $\tmmathbf{q}_{\|}$ of the
momentum separation must be zero, thus restricting a
$\tmmathbf{q}$-integration to the plane perpendicular to the momentum change
$\tmmathbf{p}_f -\tmmathbf{p}_i$,
\begin{eqnarray}
  \xi \left( \tmmathbf{p}_f, \tmmathbf{p}_i ; \tmmathbf{q} \right) & = &
  \left\{ \begin{array}{ll}
    \xi \left( \tmmathbf{p}_f, \tmmathbf{p}_i ; \tmmathbf{q}_{\bot} \right) &
    \text{if $\tmmathbf{q}_{\|} = 0$}\\
    0 & \text{otherwise} .
  \end{array} \right. 
\end{eqnarray}
The formal square root of the product of delta functions then reduces to a
single proper Dirac function $\delta \left( \left( \tmmathbf{p}_f^2
-\tmmathbf{p}_i^2 \right) / 2 \right)$, and we obtain, as the natural
generalization of (\ref{eq:replacementrule}),
\begin{align}
  \xi \left( \tmmathbf{p}_f, \tmmathbf{p}_i ; \tmmathbf{q}_{\bot} \right)  =
  & \delta \left( \frac{\tmmathbf{p}_f^2 -\tmmathbf{p}_i^2}{2} \right) 
  \frac{f \left( \tmmathbf{p}_f +\tmmathbf{q}_{\bot}, \tmmathbf{p_i}
  +\tmmathbf{q}_{\bot} \right)}{\sqrt{\sigma_{\tmop{tot}} \left(
  \tmmathbf{p_i} +\tmmathbf{q}_{\bot} \right)  \left| \tmmathbf{p_i}
  +\tmmathbf{q}_{\bot} \right|}}  \nonumber\\
    & \times \frac{f^{\ast} \left( \tmmathbf{p}_f -\tmmathbf{q}_{\bot},
  \tmmathbf{p_i} -\tmmathbf{q}_{\bot} \right)}{\sqrt{\sigma_{\tmop{tot}}
  \left( \tmmathbf{p_i} -\tmmathbf{q}_{\bot} \right)  \left| \tmmathbf{p_i}
  -\tmmathbf{q}_{\bot} \right|}} .  \label{eq:grr}
\end{align}
Inserting this expression into (\ref{eq:minxi}), together with rates
determined by (\ref{eq:Gamzero}), one arrives directly at the complex rate
density $m_{\tmop{in}}$ given by Eqs.~(\ref{eq:mclaim0}), (\ref{eq:mclaim}).

We emphasize again that, compared to the microscopic phase space description
in the preceding section, the line of reasoning is here quite different, and
indeed more heuristic. The fact that the two lines of argument yield identical
results indicates that their specific assumptions do reflect the underlying
physics.

\section{The gas induced energy shift }\label{sec:refrac}

We now turn to the second part of the master equation, given by the
superoperator $\mathcal{R}$ defined in Eq.~(\ref{eq:Rcal}). This term
describes the coherent modification of the tracer particle dynamics due to the
presence of the gas. As with the incoherent part $\mathcal{L}$ given in
(\ref{eq:me1}), a naive evaluation would take the expressions (\ref{eq:Gam3})
and (\ref{eq:Smatrix}) for the rate and scattering operators at face value,
and would thus yield an ill-defined normalization involving a $\delta \left( 0
\right) / | \Omega |$ term. The correct normalization will be obtained in this
section by implementing the appropriate restriction to the incoming states in
the same way as in Sect.~\ref{sec:wp}.

It should be noted that the effect of the energy shift described by
$\mathcal{R}$ can usually be neglected when the incoherent effects of the
master equation play a role so that $\mathcal{L}$ dominates the dynamics.
However, one can set up atom interferometer experiments where one beam
interacts with a gas filled region such that only those atoms contribute to
the detected signal which did not change their momentum by a collision. In
this case, the effect can be measured as a gas-induced phase shift, and it is
usually accounted for by attributing a refractive index to the gas
{\cite{Schmiedmayer1995a,Vigue1995a,Forrey1996a,Kharchenko2001a,Jacquey2007a}}.

Before starting the calculation, let us note that in
Ref.~{\cite{Hornberger2007b}} a slightly different term was given for the
coherent modification, namely $\mathcal{R}' \rho = \mathi
\tmop{Tr}_{\tmop{gas}} ([\tmop{Re} \left( \mathsf{T} \right),
\mathsf{\Gamma}^{1 / 2} \left[ \rho \otimes \rho_{\tmop{gas}} \right]
\mathsf{\Gamma}^{1 / 2}])$. It differs from (\ref{eq:Rcal}) in the location of
one of the $\mathsf{\Gamma}^{1 / 2}$ operators. In fact, the two
superoperators $\mathcal{R}$ and $\mathcal{R}'$ yield the same gas induced
energy shift when applied to the immobile system discussed in
{\cite{Hornberger2007b}}. For the present case of a tracer particle,
$\mathcal{R}'$ has the disadvantage of introducing a weak dependence on $P /
P'$ which would need to be removed as an additional approximation. It
therefore seems more natural to start from the form (\ref{eq:Rcal}) right
away, which is manifestly unitary.

The calculational procedure can be carried out in complete analogy to the
reasoning in Sect.~\ref{sec:wp}. Inserting the stationary state
(\ref{eq:rhogasmu}) into the expression (\ref{eq:Rcal}) for $\mathcal{R} \rho$
yields immediately the coherent modification of the evolution due to the
presence of the gas. In momentum representation it takes the form of
Eq.~(\ref{eq:cohexp1}) with the energy shifts given by
\begin{eqnarray}
  E_{\text{n}} \left( \tmmathbf{P} \right) & = & - \hbar \frac{\left( 2 \pi
  \hbar \right)^3}{| \Omega |} \int \mathd \tmmathbf{p}_0 \, \mu \left(
  \tmmathbf{p}_0 \right)   \label{eq:Endef}\\
  &  & \times \langle \tmop{rel} \left( \tmmathbf{p}_0, \tmmathbf{P} \right)
  | \mathsf{\Gamma}_0^{1 / 2} \tmop{Re} \left( \mathsf{T}_0 \right)
  \mathsf{\Gamma}_0^{1 / 2} | \tmop{rel} \left( \tmmathbf{p}_0, \tmmathbf{P}
  \right) \rangle . \nonumber
\end{eqnarray}
We can again switch to the center-of-mass frame by introducing the relative
momentum
\begin{eqnarray}
  \tmmathbf{p}_n & = & \tmop{rel} \left( \tmmathbf{p}_0, \tmmathbf{P} \right) 
\end{eqnarray}
as a function of $\tmmathbf{p}_0$. This way the energy shifts take the form of
an average over the gas momentum distribution function $\mu$,
\begin{eqnarray}
  E_{\text{n}} \left( \tmmathbf{P} \right) & = & \int \mathd \tmmathbf{p}_0 \,
  \mu \left( \tmmathbf{p}_0 \right) e_{\text{n}} \left( \tmmathbf{p}_n \right)
  .  \label{eq:Endef2}
\end{eqnarray}
Like in the incoherent case, the function to be averaged can again be written
as an expectation value with respect to a normalized momentum state of the
relative motion $\rho_{\tmmathbf{p}_n} =\|\tmmathbf{p}_n \rangle \rangle
\langle \langle \tmmathbf{p}_n ||$. The function has the unit of an energy,
\begin{eqnarray}
  e_{\text{n}} \left( \tmmathbf{p}_n \right) & = & - \hbar \tmop{Tr} \left(
  \mathsf{\Gamma}_0^{1 / 2} \tmop{Re} \left( \mathsf{T}_0 \right)
  \mathsf{\Gamma}_0^{1 / 2} \rho_{\tmmathbf{p}_n} \right) .  \label{eq:endef}
\end{eqnarray}
When evaluating the expectation value, the restriction to incoming wave
packets can again be implemented by replacing $\rho_{\tmmathbf{p}_n}$ with its
restricted version $\rho_{\tmmathbf{p}_n}'$. It is given by
Eq.~(\ref{eq:rhopip}) with $\tmmathbf{p}_i$ replaced by $\tmmathbf{p}_n$, see
Fig.~\ref{fig:cylinder}. One thus obtains
\begin{align}
  e_{\text{n}} \left( \tmmathbf{p}_n \right)  = & - \frac{1}{2 \pi} \int
  \mathd \tmmathbf{w} \int_{\Lambda_{\tmmathbf{p}_n}} \frac{\mathd
  \tmmathbf{x}_{\|\tmmathbf{p}_n}}{| \Lambda_{\tmmathbf{p}_n} |}
  \int_{\Sigma_{\tmmathbf{p}_n}} \frac{\mathd \tmmathbf{x}_{\bot
  \tmmathbf{p}_n}}{| \Sigma_{\tmmathbf{p}_n} |} \bignone  \nonumber\\
    & \times \exp \left( i \frac{\tmmathbf{x} \cdot \tmmathbf{w}}{\hbar}
  \right) \Gamma_0^{1 / 2} \left( \tmmathbf{p}_n + \frac{\tmmathbf{w}}{2}
  \right) \Gamma_0^{1 / 2} \left( \tmmathbf{p}_n - \frac{\tmmathbf{w}}{2}
  \right) \nonumber\\
    & \times \delta \left( \tmmathbf{p}_n \cdot \tmmathbf{w} \right)
  \tmop{Re} \left[ f \left( \tmmathbf{p}_n + \frac{\tmmathbf{w}}{2},
  \tmmathbf{p}_n - \frac{\tmmathbf{w}}{2} \right) \right] . \nonumber
\end{align}
Like in Sect.~\ref{sec:wp}, we identify the cylinder base area with the
scattering cross section, $| \Sigma_{\tmmathbf{p}_n} | = \sigma \left(
\tmmathbf{p}_n \right)$, and we note that the delta-function removes the
component of $\tmmathbf{w}$ which is parallel to $\tmmathbf{p}_n$, so that the
dependence on $\tmmathbf{x}_{\|\tmmathbf{p}_n}$ vanishes in the integrand,
\begin{eqnarray}
  e_{\text{n}} \left( \tmmathbf{p}_n \right) & = & - 2 \pi \hbar^2
  \frac{n_{\tmop{gas}}}{m_{\ast}} \int \mathd \tmmathbf{w} \, \delta \left(
  \frac{\tmmathbf{p}_n \cdot \tmmathbf{w}}{p_n} \right) \nonumber\\
  &  & \times \int_{\Sigma_{\tmmathbf{p}_n}} \frac{\mathd \tmmathbf{x}_{\bot
  \tmmathbf{p}_n}}{\left( 2 \pi \hbar \right)^2} \bignone \exp \left( i
  \frac{\tmmathbf{x}_{\bot \tmmathbf{p}_n} \cdot \tmmathbf{w}_{\bot
  \tmmathbf{p}_n}}{\hbar} \right)  \nonumber\\
  &  & \times \int_{\Lambda_{\tmmathbf{p}_n}} \frac{\mathd
  \tmmathbf{x}_{\|\tmmathbf{p}_n}}{| \Lambda_{\tmmathbf{p}_n} |} \tmop{Re}
  \left[ f \left( \tmmathbf{p}_n + \frac{\tmmathbf{w}}{2}, \tmmathbf{p}_n -
  \frac{\tmmathbf{w}}{2} \right) \right] \nonumber\\
  &  & \times \frac{\sqrt{|\tmmathbf{p}_n + \frac{\tmmathbf{w}}{2} | \,
  |\tmmathbf{p}_n - \frac{\tmmathbf{w}}{2} |}}{p_n} \nonumber\\
  &  & \times \frac{\sqrt{\sigma \left( \tmmathbf{p}_n +
  \frac{\tmmathbf{w}}{2} \right) \sigma \left( \tmmathbf{p}_n -
  \frac{\tmmathbf{w}}{2} \right)}}{\sigma \left( \tmmathbf{p}_n \right)} .
  \nonumber
\end{eqnarray}
Carrying out the $\tmmathbf{x}_{\|\tmmathbf{p}_n}$-integration,
$\int_{\Lambda_{\tmmathbf{p}_n}} \mathd \tmmathbf{x}_{\|\tmmathbf{p}_n} = |
\Lambda_{\tmmathbf{p}_n} |$, one observes that the remaining
$\tmmathbf{x}_{\bot \tmmathbf{p}_n}$-integration yields an approximate
two-dimensional delta-function in $\tmmathbf{w}_{\bot \tmmathbf{p}_n}$.
Combined with the delta function in $\tmmathbf{p}_n \cdot \tmmathbf{w}/ p_n =
w_{\|\tmmathbf{p}_n}$ this gives a three-dimensional $\delta \left(
\tmmathbf{w} \right)$, which permits to do the $\tmmathbf{w}$-integration. One
thus obtains
\begin{eqnarray}
  e_{\text{n}} \left( \tmmathbf{p}_n \right) & = & - 2 \pi \hbar^2
  \frac{n_{\tmop{gas}}}{m_{\ast}} \tmop{Re} \left[ f \left( \tmmathbf{p}_n,
  \tmmathbf{p}_n \right) \right] .  \label{eq:endef2}
\end{eqnarray}
It shows that the energy shift is essentially determined by the real part of
the forward scattering amplitude, a fact that is well-known in the field of
neutron and atom optics. Its effect is often expressed by introducing an index
of refraction $n _1$, as discussed in Sect.~\ref{sec:ior}.

\section{Operator representation of the quantum linear Boltzmann
equation}\label{sec:or}

So far, the derivation of the master equation was discussed in the momentum
representation. Let us now turn to the question how to obtain the quantum
linear Boltzmann equation in a representation-independent form. The result
will then immediately prove the complete positivity and the translational
covariance of the dynamical map defined by the master equation.

The calculations in Sects.~\ref{sec:wp} and \ref{sec:rr} both indicate that
$m_{\tmop{in}}$, the rate function in the center of mass frame, is given by
Eq.~(\ref{eq:mclaim}). The complex rate $M_{\tmop{in}}$, which determines the
incoherent evolution in momentum representation according to
(\ref{eq:Dtrho1}), is then obtained by averaging $m_{\tmop{in}}$ with the gas
momentum distribution function $\mu$. Specifically, Eq.~(\ref{eq:Minrel})
tells that $M_{\tmop{in}} \left( \tmmathbf{P}, \tmmathbf{P}' ; \tmmathbf{Q}
\right) = \int \mathd \tmmathbf{p}_0 \mu \left( \tmmathbf{p}_0 \right)
m_{\tmop{in}} \left( \tmmathbf{p}_f, \tmmathbf{p}_i ; \tmmathbf{q} \right)$
with the relative momenta $\tmmathbf{p}_f$, $\tmmathbf{p}_i$, and
$\tmmathbf{q}$ defined in (\ref{eq:pidef})-(\ref{eq:qdef}). However, the
resulting expression is not of the factorized form (\ref{eq:Min3}) needed
below when stating the master equation in its operator representation.

To arrive at (\ref{eq:Min3}) we first change the integration variable from
$\tmmathbf{p}_0$ to $\tmmathbf{p}_i$. Moreover, the $\mu$ distribution can be
split symmetrically into a product of square roots, $\mu \left( \tmmathbf{p}_0
\right) = \mu^{1 / 2} \left( \tmmathbf{p}_0 \right) \mu^{1 / 2} \left(
\tmmathbf{p}_0 \right)$, since $\tmmathbf{p}_0$ can be equally expressed as
$\tmmathbf{p}_i + \left( \tmmathbf{p}_f +\tmmathbf{P} \right) m / M
+\tmmathbf{q}m / m_{\ast}$ or as $\tmmathbf{p}_i + \left( \tmmathbf{p}_f
+\tmmathbf{P}' \right) m / M -\tmmathbf{q}m / m_{\ast}$, see
(\ref{eq:pidef})-(\ref{eq:qdef}). Noting $| \det \left( \partial
\tmmathbf{p}_0 / \partial \tmmathbf{p}_i \right) | = m^3 / m^3_{\ast}$ we have
\begin{eqnarray}
  &  & M_{\tmop{in}} \left( \tmmathbf{P}, \tmmathbf{P}' ; \tmmathbf{Q}
  \right) \nonumber\\
  &  & = \hspace{0.6em} \left( \frac{m}{m_{\ast}} \right)^3 \int \mathd
  \tmmathbf{p}_i \, \mu^{1 / 2} \left( \tmmathbf{p}_i + \frac{m}{M}  \left(
  \tmmathbf{p}_f +\tmmathbf{P} \right) + \frac{m}{m_{\ast}}
  \tmmathbf{q}_{\bot} \right)  \nonumber\\
  &  & \phantom{= \hspace{0.6em}} \times \mu^{1 / 2} \left( \tmmathbf{p}_i +
  \frac{m}{M}  \left( \tmmathbf{p}_f +\tmmathbf{P}' \right) -
  \frac{m}{m_{\ast}} \tmmathbf{q}_{\bot} \right) \nonumber\\
  &  & \phantom{= \hspace{0.6em}} \times m_{\tmop{in}} \left( \tmmathbf{p}_f,
  \tmmathbf{p}_i ; \tmmathbf{q}_{\bot} \right), \nonumber
\end{eqnarray}
where we replaced $\tmmathbf{q}$ by $\tmmathbf{q}_{\bot}$ in the arguments of
$\mu^{1 / 2}$, in accordance with Eq.~(\ref{eq:mrelprop}). Having implemented
the property (\ref{eq:mclaim0}) of the generalized function $m_{\tmop{in}}$,
we can now insert its explicit form (\ref{eq:mclaim}), which introduces the
scattering amplitudes and an energy conserving delta-function,
\begin{align}
    & M_{\tmop{in}} \left( \tmmathbf{P}, \tmmathbf{P}' ; \tmmathbf{Q}
  \right) \nonumber\\
    & = \hspace{0.6em} \left( \frac{m}{m_{\ast}} \right)^3 \int \mathd
  \tmmathbf{p}_i \, \mu^{1 / 2} \left( \tmmathbf{p}_i + \frac{m}{M}  \left(
  \tmmathbf{p}_f +\tmmathbf{P} \right) + \frac{m}{m_{\ast}}
  \tmmathbf{q}_{\bot} \right)  \nonumber\\
    & \phantom{= \hspace{0.6em}} \times \mu^{1 / 2} \left( \tmmathbf{p}_i +
  \frac{m}{M}  \left( \tmmathbf{p}_f +\tmmathbf{P}' \right) -
  \frac{m}{m_{\ast}} \tmmathbf{q}_{\bot} \right) \delta \left(
  \frac{\tmmathbf{p}_f^2 -\tmmathbf{p}_i^2}{2} \right) \nonumber\\
    & \phantom{= \hspace{0.6em}} \times \frac{n_{\tmop{gas}}}{m_{\ast}} f
  \left( \tmmathbf{p}_f +\tmmathbf{q}_{\bot}, \tmmathbf{p}_i
  +\tmmathbf{q}_{\bot} \right) f^{\ast} \left( \tmmathbf{p}_f
  -\tmmathbf{q}_{\bot}, \tmmathbf{p}_i -\tmmathbf{q}_{\bot} \right) .
  \nonumber\\
    & 
\end{align}
From here it is a small step to arrive at the explicit expression given in
Eq.~(\ref{eq:Mexp}). In order to obtain a factorized expression we rather
perform another change of variables,
\begin{eqnarray}
  \tmmathbf{p}_i & \rightarrow & \tmmathbf{p}= \frac{m}{m_{\ast}}
  \tmmathbf{p}_i + \frac{m}{M}  \frac{\tmmathbf{P}_{\bot \tmmathbf{Q}}
  +\tmmathbf{P}_{\bot \tmmathbf{Q}}'}{2} - \frac{m}{m_{\ast}} 
  \frac{\tmmathbf{Q}^{}}{2} . \nonumber
\end{eqnarray}
Due to its dependence on the transverse $\tmmathbf{P}$ and $\tmmathbf{P}'$
components this transformation has the remarkable effect of producing an
integrand which is a product of $\tmmathbf{P}$- and $\tmmathbf{P}'$-dependent
factors:
\begin{align}
    & M_{\tmop{in}} \left( \tmmathbf{P}, \tmmathbf{P}' ; \tmmathbf{Q}
  \right) \nonumber\\
    & \hspace{0.6em} = \hspace{0.6em} \frac{n_{\tmop{gas}}}{m_{\ast}}  \int
  \mathd \tmmathbf{p} \, \mu^{1 / 2} \left( \tmmathbf{p}+ \frac{m}{M}
  \tmmathbf{P}_{\|} + \left( 1 - \frac{m}{M} \right) \frac{\tmmathbf{Q}}{2}
  \right)  \nonumber\\
    & \phantom{\hspace{0.6em} =} \times \mu^{1 / 2} \left( \tmmathbf{p}+
  \frac{m}{M} \tmmathbf{P}'_{\|} + \left( 1 - \frac{m}{M} \right)
  \frac{\tmmathbf{Q}}{2} \right) \, \delta \left( \frac{m_{\ast}}{m}
  \tmmathbf{p} \cdot \tmmathbf{Q} \right) \nonumber\\
    & \phantom{\hspace{0.6em} =} \times f \left( \tmop{rel} \left(
  \tmmathbf{p}, \tmmathbf{P}_{\bot} \right) - \frac{\tmmathbf{Q}}{2},
  \tmop{rel} \left( \tmmathbf{p}, \tmmathbf{P}_{\bot} \right) +
  \frac{\tmmathbf{Q}}{2} \right) \nonumber\\
    & \phantom{\hspace{0.6em} =} \times f^{\ast} \left( \tmop{rel} \left(
  \tmmathbf{p}, \tmmathbf{P}_{\bot}' \right) - \frac{\tmmathbf{Q}}{2},
  \tmop{rel} \left( \tmmathbf{p}, \tmmathbf{P}_{\bot}' \right) +
  \frac{\tmmathbf{Q}}{2} \right) \nonumber\\
    & = \int \mathd \tmmathbf{p} \, \delta \left(  \frac{\tmmathbf{p} \cdot
  \tmmathbf{Q}}{Q} \right) L \left( \tmmathbf{p}, \tmmathbf{P}-\tmmathbf{Q};
  \tmmathbf{Q} \right) L^{\ast} \left( \tmmathbf{p}, \tmmathbf{P}'
  -\tmmathbf{Q}; \tmmathbf{Q} \right) . \nonumber\\
    & 
\end{align}
The second equality, which brings about the functions $L \left( \tmmathbf{p},
\tmmathbf{P}; \tmmathbf{Q} \right)$ defined in Eq.~(\ref{eq:Ldef}), emphasizes
the factorization. Observing that the delta function restricts the
$\tmmathbf{p}$-integration to the plane $\tmmathbf{Q}^{\bot} = \left\{
\tmmathbf{p} \in \mathbbm{R}^3 : \tmmathbf{p} \cdot \tmmathbf{Q}= 0 \right\}$
perpendicular to $\tmmathbf{Q}$ finally leads to the expression announced in
Eq. (\ref{eq:Min3}), since for any function $g \left( \tmmathbf{p} \right)$
\begin{eqnarray}
  \int \mathd \tmmathbf{p} \, \, \delta \left(  \frac{\tmmathbf{p} \cdot
  \tmmathbf{Q}}{Q} \right) g \left( \tmmathbf{p} \right) & = &
  \int_{\tmmathbf{Q}^{\bot}} \mathd \tmmathbf{p} \, g \left( \tmmathbf{p}
  \right) . 
\end{eqnarray}

The function $L$ from Eq.~(\ref{eq:Ldef}) is clearly well-suited to
characterize the master equation, since it contains all the details of the
collisional interaction with the gas. It comprises the elastic scattering
amplitude $f \left( \tmmathbf{p}_f, \tmmathbf{p}_i \right)$ defined by the
two-body interaction, the mass $M$ of the tracer particle, the momentum
distribution function $\mu \left( \tmmathbf{p} \right)$ of the gas, its mass
$m$ and number density $n_{\tmop{gas}}$. Unsurprisingly, the function $L$
plays a central role for the operator representation of the master equation as
well. It permits to define a family of jump operators acting in the Hilbert
space of the tracer particle,
\begin{eqnarray}
  \mathsf{L} _{\tmmathbf{Q}, \tmmathbf{p}} & = & \mathe^{i \mathsf{X \cdot
  \tmmathbf{Q}/ \hbar}} L \left( \tmmathbf{p}, \mathsf{P} ; \tmmathbf{Q}
  \right),  \label{eq:Lopdef}
\end{eqnarray}
where $\mathsf{X}$ and $\mathsf{P}$ are the corresponding position and
momentum operators (and the function $L$ is given in Eq.~(\ref{eq:Ldef})). The
first factor effects a momentum exchange by $\tmmathbf{Q}$, since $\exp \left(
i \mathsf{X \cdot \tmmathbf{Q}/ \hbar} \right) |\tmmathbf{P} \rangle =
|\tmmathbf{P}+\tmmathbf{Q} \rangle$, while the appearance of $\mathsf{P}$ in
the second factor renders the function $L$ operator valued. This implies that
both the scattering amplitude and the momentum distribution function attain an
operator character in (\ref{eq:Ldef}), which is possible because the
$\tmmathbf{P}$-dependence of $L$ will be analytic for any physically
reasonable interaction potential.

With the jump operators (\ref{eq:Lopdef}) at hand it is straightforward to
construct the superoperator $\mathcal{L}$, whose momentum representation is
given by Eq.~(\ref{eq:Dtrho1}) with $M_{\tmop{in}}$ from (\ref{eq:Min3}),
\begin{eqnarray}
  \mathcal{L} \rho & = & \bigintlim \mathd \tmmathbf{Q}
  \int_{\tmmathbf{Q}^{\bot}} \mathd \tmmathbf{p} \left\{ \mathsf{L}
  _{\tmmathbf{Q}, \tmmathbf{p}} \rho \mathsf{L} _{\tmmathbf{Q},
  \tmmathbf{p}}^{\dag} - \frac{1}{2} \rho \mathsf{L} _{\tmmathbf{Q},
  \tmmathbf{p}}^{\dag} \mathsf{L} _{\tmmathbf{Q}, \tmmathbf{p}} \right.
  \nonumber\\
  &  & \left. - \frac{1}{2}  \mathsf{L} _{\tmmathbf{Q}, \tmmathbf{p}}^{\dag}
  \mathsf{L} _{\tmmathbf{Q}, \tmmathbf{p}} \rho \right\} \bignone . 
  \label{eq:qlbe}
\end{eqnarray}
This is the equation given in Ref.~{\cite{Hornberger2006b}} (up to a trivial
change of notation).

It is reassuring to observe that the form of the generator (\ref{eq:qlbe}) is
in accordance with the most general structure of a translation-invariant and
completely positive master equation as characterized by Holevo
{\cite{Holevo1996a}}, see {\cite{Petruccione2005a,Vacchini2005a}} for a
discussion. However, the summation in Ref.\, {\cite{Holevo1996a}} is here
replaced by the $\tmmathbf{p}$-integration over the plane
$\tmmathbf{Q}^{\bot}$ in (\ref{eq:qlbe}).

A further consistency requirement is based on the transformation to a moving
frame of reference. Denoting the velocity boost by $\tmmathbf{V}$, the
transformed state of the tracer particle is given by
\begin{eqnarray}
  \rho_{\tmmathbf{V}} & = & \mathe^{i \mathsf{X} \cdot M\tmmathbf{V}/ \hbar}
  \rho \mathe^{- i \mathsf{X} \cdot M\tmmathbf{V}/ \hbar}, \nonumber
\end{eqnarray}
and the incoherent evolution in the new frame of reference
$\mathcal{L}_{\tmmathbf{V}}$ is thus related to $\mathcal{L}$ by
\begin{eqnarray}
  \mathcal{L}_{\tmmathbf{V}}  \left[ \cdot \right] & = & \mathe^{i \mathsf{X}
  \cdot M\tmmathbf{V}/ \hbar} \mathcal{L} \left[ \mathe^{- i \mathsf{X} \cdot
  M\tmmathbf{V}/ \hbar} \cdot \mathe^{i \mathsf{X} \cdot M\tmmathbf{V}/ \hbar}
  \right] \mathe^{- i \mathsf{X} \cdot M\tmmathbf{V}/ \hbar} \nonumber\\
  &  &  \label{eq:Lcalv}
\end{eqnarray}
However, the same super-operator must be obtained if we {\tmem{actively}}
shift the momentum distribution $\mu \left( \tmmathbf{p} \right)$ of the
background gas, by setting $\mu_{\tmmathbf{V}} \left( \tmmathbf{p} \right) =
\mu \left( \tmmathbf{p}- m\tmmathbf{V} \right)$ in the function $L$ defining
the jump operators (\ref{eq:Lopdef}). The reason why this transformation of
the gas motion must have the same effect as (\ref{eq:Lcalv}) is that the
interaction between the tracer particle and the gas depends only on their
relative motion. Indeed, the functions $L$ and $L_{\tmmathbf{V}}$, based on
the gas distributions $\mu$ and $\mu_{\tmmathbf{V}}$ in (\ref{eq:Ldef}), are
related by $L \left( \tmmathbf{p}, \tmmathbf{P}- M\tmmathbf{V}; \tmmathbf{Q}
\right) = L_{\tmmathbf{V}} \left( \tmmathbf{p}+ m\tmmathbf{V}_{\bot
\tmmathbf{Q}}, \tmmathbf{P}; \tmmathbf{Q} \right)$. Noting also that a change
of the integration variable $\tmmathbf{p} \rightarrow \tmmathbf{p}_{\bot}'
=\tmmathbf{p}+ m\tmmathbf{V}_{\bot \tmmathbf{Q}}$ in (\ref{eq:qlbe}) is
possible, since it leaves the plane $\tmmathbf{Q}^{\bot}$ invariant, one
easily proves the equivalence of the coordinate transformation and the shift
of the momentum distribution.

As a final step, let us also incorporate the coherent modification of the
tracer dynamics as discussed in Sect.~\ref{sec:refrac}. The energy shift
operator
\begin{eqnarray}
  \mathsf{H}_{\text{n}} & = & E_{\text{n}} \left( \mathsf{P} \right) 
  \label{eq:Hndef}
\end{eqnarray}
is given by the operator-valued version of Eq.~(\ref{eq:cohexp2}). It permits
to write the coherent modification part of the master equation
(\ref{eq:cohexp1}) as $\mathcal{R} \rho = \left( i \hbar \right)^{- 1} \left[
\mathsf{H}_{\text{n}}, \rho \right] .$ This super-operator has the same
invariance and transformation properties as discussed above in the case of
$\mathcal{L}$. In particular, its transformation to a moving frame of
reference analogous to (\ref{eq:Lcalv}) is equally obtained by replacing $\mu$
with $\mu_{\tmmathbf{V}}$ in (\ref{eq:cohexp2}).

To summarize this section we include the free motion Hamiltonian $\mathsf{H} =
\mathsf{P}^2 / 2 M$, thus writing the complete quantum linear Boltzmann
equation (\ref{eq:me0}) in the representation-independent form
\begin{align}
  \partial_t \rho  = & \frac{1}{i \hbar} \left[ \frac{\mathsf{P}^2}{2 M} +
  \mathsf{H}_{\text{n}}, \rho \right] + \frac{1}{2} \bigintlim \mathd
  \tmmathbf{Q} \int_{\tmmathbf{Q}^{\bot}} \mathd \tmmathbf{p} \left\{ \left[
  \mathsf{L} _{\tmmathbf{Q}, \tmmathbf{p}}, \rho \mathsf{L} _{\tmmathbf{Q},
  \tmmathbf{p}}^{\dag} \right] \right. \nonumber\\
  &   \left. + \left[ \mathsf{L} _{\tmmathbf{Q}, \tmmathbf{p}} \rho,
  \mathsf{L} _{\tmmathbf{Q}, \tmmathbf{p}}^{\dag} \right] \right\} \bignone . 
  \label{eq:qlbe2}
\end{align}
\section{Limiting forms}\label{sec:lf}

As an important cross-check of the QLBE derived above, let us now see whether
taking suitable limits reduces its form to that of previously established
equations.

\subsection{Classical linear Boltzmann equation}

The most obvious limiting motion is that of a classical particle. If all
off-diagonal elements vanish in a motional state, $\langle \tmmathbf{P}| \rho
|\tmmathbf{P}' \rangle = 0$, it is characterized by the diagonal momentum
distribution $f_{\text{p}} \left( \tmmathbf{P} \right) = \text{$\langle
\tmmathbf{P}| \rho |\tmmathbf{P} \rangle$}$ alone, and insofar it is
indistinguishable from a classical state. One expects that the motion of the
diagonal elements predicted by (\ref{eq:qlbe}) is equal to the one described
by the classical linear Boltzmann equation.

As follows from the discussion in Sect.~\ref{sec:momemtumQLBE}, the QLBE
implies that the diagonal elements $f_{\text{p}} \left( \tmmathbf{P} \right)$
satisfy
\begin{eqnarray}
  \partial^{\tmop{coll}}_t f_{\text{p}} (\tmmathbf{P}) & = & \int \mathd
  \tmmathbf{Q} \bignone M_{\tmop{in}}^{\tmop{cl}} \left( \tmmathbf{P};
  \tmmathbf{Q} \right) f_{\text{p}} (\tmmathbf{P}-\tmmathbf{Q}) \nonumber\\
  &  & - M_{\tmop{out}}^{\tmop{cl}} \left( \tmmathbf{P} \right) f_{\text{p}}
  (\tmmathbf{P}),  \label{eq:QLBEcl}
\end{eqnarray}
with the rates $M_{\tmop{out}}^{\tmop{cl}} \left( \tmmathbf{P} \right)$ and
$\bignone M_{\tmop{in}}^{\tmop{cl}} \left( \tmmathbf{P}; \tmmathbf{Q} \right)$
given by Eqs.~(\ref{eq:Mcloutexp}) and (\ref{eq:Mclin}). The notation
$\partial^{\tmop{coll}}_t$ indicates that we focus here only on the
differential change in time which is due to the collision part $\mathcal{L}$
of the master equation.

This equation should be compared to the classical linear Boltzmann equation
{\cite{Cercignani1975a}} for the momentum distribution function
$f_{\text{p}}^{\tmop{cl}} (\tmmathbf{P})$. The traditional form of the
collision integral reads, in our notation,
\begin{eqnarray}
  \partial^{\tmop{coll}}_t f_{\text{p}}^{\tmop{cl}} (\tmmathbf{P}) & = &
  n_{\tmop{gas}} \bigintlim \mathd \tmmathbf{p} \mathd \tmmathbf{n} \frac{|
  \tmop{rel} \left( \tmmathbf{p}, \tmmathbf{P} \right) |}{m_{\ast}} 
  \nonumber\\
  &  & \times \sigma \left( \left| \tmop{rel} \left( \tmmathbf{p},
  \tmmathbf{P} \right) \right| \tmmathbf{n}, \tmop{rel} \left( \tmmathbf{p},
  \tmmathbf{P} \right) \right) \nonumber\\
  &  & \times \left\{ \mu \left( \tmmathbf{p}' \right)
  f_{\text{p}}^{\tmop{cl}} \left( \tmmathbf{P}' \right) - \mu \left(
  \tmmathbf{p} \right) f_{\text{p}}^{\tmop{cl}} \left( \tmmathbf{P} \right)
  \right\},  \nonumber\\\label{eq:CLBE}
\end{eqnarray}
where $\tmmathbf{n}$ is the unit vector of an angular integration with $\mathd
\tmmathbf{n}$ the associated element of solid angle. The values of
$\tmmathbf{P}'$ and $\tmmathbf{p}'$ are determined by momentum conservation,
granting in particular $| \tmop{rel} \left( \tmmathbf{p}', \tmmathbf{P}'
\right) | = | \tmop{rel} \left( \tmmathbf{p}, \tmmathbf{P} \right) |$. Using
the PT-invariance of the differential cross section, $\sigma \left(
\tmmathbf{p}_f, \tmmathbf{p}_i \right) = \sigma \left( \tmmathbf{p}_i,
\tmmathbf{p}_f \right)$, the classical linear Boltzmann equation can thus be
rewritten in the explicit form
\begin{eqnarray}
  \partial^{\tmop{coll}}_t f_{\text{p}}^{\tmop{cl}} (\tmmathbf{P}) & = &
  \frac{n_{\tmop{gas}}}{m_{\ast}} \bigintlim \mathd \tmmathbf{p} \mathd
  \tmmathbf{n}_i | \tmop{rel} \left( \tmmathbf{p}, \tmmathbf{P} \right) | 
  \label{eq:87}\\
  &  & \times \mu \left( \tmmathbf{p}- \tmop{rel} \left( \tmmathbf{p},
  \tmmathbf{P} \right) + \left| \tmop{rel} \left( \tmmathbf{p}, \tmmathbf{P}
  \right) \right| \tmmathbf{n}_i \right)  \nonumber\\
  &  & \times \sigma \left( \tmop{rel} \left( \tmmathbf{p}, \tmmathbf{P}
  \right), \left| \tmop{rel} \left( \tmmathbf{p}, \tmmathbf{P} \right) \right|
  \tmmathbf{n}_i \right)  \nonumber\\
  &  & \times f_{\text{p}}^{\tmop{cl}} \left( \tmmathbf{P}+ \tmop{rel} \left(
  \tmmathbf{p}, \tmmathbf{P} \right) - \left| \tmop{rel} \left( \tmmathbf{p},
  \tmmathbf{P} \right) \right| \tmmathbf{n}_i \right) \nonumber\\
  &  & - f_{\text{p}}^{\tmop{cl}} \left( \tmmathbf{P} \right) 
  \frac{n_{\tmop{gas}}}{m_{\ast}} \bigintlim \mathd \tmmathbf{p} \mathd
  \tmmathbf{n}_f | \tmop{rel} \left( \tmmathbf{p}, \tmmathbf{P} \right) |
  \nonumber\\
  &  & \times \mu \left( \tmmathbf{p} \right) \sigma \left( \left| \tmop{rel}
  \left( \tmmathbf{p}, \tmmathbf{P} \right) \right| \tmmathbf{n}_f, \tmop{rel}
  \left( \tmmathbf{p}, \tmmathbf{P} \right) \right) . \nonumber
\end{eqnarray}
The angular integrations can be converted into three-dimensional integrals
with a delta function. Noting
\begin{eqnarray}
  &  & \left| \tmop{rel} \left( \tmmathbf{p}, \tmmathbf{P} \right) \right|
  \delta \left( |\tmmathbf{p}_{i, f} | - \left| \tmop{rel} \left(
  \tmmathbf{p}, \tmmathbf{P} \right) \right| \right) \nonumber\\
  &  & = \hspace{0.6em} p_{i, f}^2 \delta \left( \frac{|\tmmathbf{p}_{i, f}
  |^2 - \left| \tmop{rel} \left( \tmmathbf{p}, \tmmathbf{P} \right)
  \right|^2}{2} \right) \nonumber
\end{eqnarray}
one arrives, after the substitutions $\tmmathbf{p}_{i, f} \rightarrow
\tmmathbf{P}_{i, f} =\tmmathbf{P}+ \tmop{rel} \left( \tmmathbf{p},
\tmmathbf{P} \right) -\tmmathbf{p}_{i, f}$, at the form
\begin{eqnarray}
  \partial^{\tmop{coll}}_t f^{\tmop{cl}}_{\text{p}} (\tmmathbf{P}) & = & \int
  \mathd \tmmathbf{P}_i \, \bignone M^{\tmop{cl}} \left( \tmmathbf{P}_i
  \rightarrow \tmmathbf{P} \right) f^{\tmop{cl}}_{\text{p}} \left(
  \tmmathbf{P}_i \right)  \label{eq:CLBE3} \nonumber\\
  &  & - f^{\tmop{cl}}_{\text{p}} \left( \tmmathbf{P} \right) \int \mathd
  \bignone \tmmathbf{P}_f \, M^{\tmop{cl}} \left( \tmmathbf{P}_{} \rightarrow
  \tmmathbf{P}_f \right) \nonumber
\\
\end{eqnarray}
with the classical rate density for the change of the tracer particle momentum
from $\tmmathbf{P}_i$ to $\tmmathbf{P}_f$ given by
\begin{align}
    & M^{\tmop{cl}} \left( \tmmathbf{P}_i \rightarrow \tmmathbf{P}_f \right)
  \nonumber\\
    & = \frac{n_{\tmop{gas}}}{m_{\ast}} \bigintlim \mathd \tmmathbf{p}_0 \,
  \mu \left( \tmmathbf{p}_0 \right) \sigma \left( \tmop{rel} \left(
  \tmmathbf{p}_0, \tmmathbf{P}_i \right) +\tmmathbf{P}_i -\tmmathbf{P}_f,
  \tmop{rel} \left( \tmmathbf{p}_0, \tmmathbf{P}_i \right) \right) \nonumber\\
    & \times \delta \left( \frac{\left| \tmop{rel} \left( \tmmathbf{p}_0,
  \tmmathbf{P}_i \right) +\tmmathbf{P}_i -\tmmathbf{P}_f \right|^2 - \left|
  \tmop{rel} \left( \tmmathbf{p}_0, \tmmathbf{P}_i) \right|^2 \right.}{2}
  \right) . 
\end{align}
It is now easy to see that the form (\ref{eq:CLBE3}) of the classical linear
Boltzmann equation is indeed identical to the diagonal part (\ref{eq:QLBEcl})
of the QLBE, with $M_{\tmop{out}}^{\tmop{cl}} \left( \tmmathbf{P} \right)$ and
$\bignone M_{\tmop{in}}^{\tmop{cl}} \left( \tmmathbf{P}; \tmmathbf{Q} \right)$
given by Eqs.~(\ref{eq:Mcloutexp}) and (\ref{eq:Mclin}).

\subsection{Pure collisional decoherence}

Another possible effect of the gas on a quantum tracer particle, and in a
sense the other extreme compared to the classical dynamics on the diagonal, is
the appearance of {\tmem{pure}} collisional decoherence. It follows from the
QLBE (\ref{eq:qlbe}) in the limit where the mass $M$ of the tracer particle is
much larger than the mass $m$ of the gas molecules, so that there is no energy
exchange during a collision. Taking ${m / M}$ to zero simplifies the function
(\ref{eq:Ldef}) characterizing the jump operators in (\ref{eq:Lopdef}), and
renders it independent of $\tmmathbf{P},$
\begin{eqnarray}
  L \left( \tmmathbf{p}, \tmmathbf{P}; \tmmathbf{Q} \right) &
  \overset{\frac{m}{M} \rightarrow 0}{\longrightarrow} &
  \sqrt{\frac{n_{\tmop{gas}} }{Qm}} \mu \left( \tmmathbf{p}_{\bot
  \tmmathbf{Q}} + \frac{\tmmathbf{Q}}{2} \right)^{1 / 2} \nonumber\\
  &  & \times f \left( \tmmathbf{p}_{\bot \tmmathbf{Q}} -
  \frac{\tmmathbf{Q}}{2}, \tmmathbf{p}_{\bot \tmmathbf{Q}} +
  \frac{\tmmathbf{Q}}{2} \right) .  \label{eq:Llim}
\end{eqnarray}
It follows that the generator of the incoherent evolution (\ref{eq:qlbe})
reduces to the form{\footnote{In the same limit $m / M \rightarrow 0$ the
energy shift operator $\mathsf{H}_{\text{n}}$ from (\ref{eq:Hndef}) turns into
a constant, so that it has no observable consequences for a constant gas
density.}}
\begin{eqnarray}
  \mathcal{L} \rho & \overset{\frac{m}{M} \rightarrow 0}{\longrightarrow} & 
  \frac{n_{\tmop{gas}}}{m} \bigintlim \mathd \tmmathbf{p}_i \mathd
  \tmmathbf{p}_f \, \mu \left( \tmmathbf{p}_i \right) \delta \left(
  \frac{\tmmathbf{p}_f^2 -\tmmathbf{p}_i^2}{2} \right) \sigma \left(
  \tmmathbf{p}_f, \tmmathbf{p}_i \right) \nonumber\\
  &  & \times \left\{ \mathe^{i \mathsf{X} \cdot \left( \tmmathbf{p}_i
  -\tmmathbf{p}_f \right) / \hbar} \rho \mathe^{- i \mathsf{X} \cdot \left(
  \tmmathbf{p}_i -\tmmathbf{p}_f \right) / \hbar} - \rho \right\} . 
  \label{eq:qlbelim}
\end{eqnarray}
This is the master equation of pure collisional decoherence discussed by
Gallis and Fleming {\cite{Gallis1990a}} and derived in its final form in
Ref.~{\cite{Hornberger2003b}}. It describes an exponential decay of the
off-diagonal elements in position representation,
\begin{eqnarray}
  \langle \tmmathbf{X}| \mathcal{L} \rho |\tmmathbf{X}' \rangle &
  \overset{\frac{m}{M} \rightarrow 0}{\longrightarrow} & - F \left(
  \tmmathbf{X}-\tmmathbf{X}' \right) \langle \tmmathbf{X}| \rho |\tmmathbf{X}'
  \rangle, 
\end{eqnarray}
with a localization rate given by
\begin{eqnarray}
  F \left( \tmmathbf{R}-\tmmathbf{R}' \right) & = & \frac{n_{\tmop{gas}}}{m}
  \bigintlim \mathd \tmmathbf{p}_i \mathd \tmmathbf{p}_f \, \mu \left(
  \tmmathbf{p}_i \right) \delta \left( \frac{\tmmathbf{p}_i^2
  -\tmmathbf{p}_f^2}{2} \right) \bignone  \nonumber\\
  &  & \times \sigma \left( \tmmathbf{p}_f, \tmmathbf{p}_i \right)  \left\{ 1
  - \mathe^{i \left( \tmmathbf{R}-\tmmathbf{R}' \right) \cdot \left(
  \tmmathbf{p}_i -\tmmathbf{p}_f \right) / \hbar} \right\} . \nonumber\\
  &  & 
\end{eqnarray}
This loss of coherence in the position basis can be attributed to the amount
of position information (or `which path' information) gained by the colliding
gas. Recently, it has been observed that interfering fullerene molecules
display a reduction of interference visibility in agreement with this equation
{\cite{Hornberger2003a,Hornberger2004a,Hackermuller2003b}}.

\subsection{Specialization to the Maxwell-Boltzmann distribution}

So far, we kept the momentum distribution $\mu$ of the gas molecules
unspecified. This served to highlight the generality of the equations and it
permitted, at the end of Sect.~\ref{sec:or}, to discuss the implications of a
transformation of the momentum distribution. However, the most important
choice is of course that of a Maxwell gas, characterized by a temperature $T =
1 / \beta k_{\text{B}}$. The remaining discussions of limiting forms in this
section will be done with the corresponding Maxwell-Boltzmann distribution,
\begin{eqnarray}
  \mu_{\beta} \left( \tmmathbf{p} \right) & = & \frac{1}{\pi^{3 / 2}
  p_{\beta}^3} \exp \left( - \frac{\tmmathbf{p}^2}{p_{\beta}^2} \right), 
  \label{eq:muMB}
\end{eqnarray}
where $p_{\beta} = \sqrt{2 m / \beta}$ is the most probable momentum. We note
that the statistical operator of the gas then takes the form
\begin{eqnarray}
  \rho_{\tmop{gas}}^{\beta} & = & \frac{\lambda_{\tmop{th}}^3}{\left| \Omega
  \right|}  \mathsf{I}_{\Omega} \exp \left( - \beta \frac{\mathsf{p}^2}{2 m}
  \right) \mathsf{I}_{\Omega}  \label{eq:rhogasMB}
\end{eqnarray}
with $\lambda_{\tmop{th}} = \sqrt{2 \pi \hbar^2 \beta / m}$ the thermal de
Broglie wave length and $\mathsf{I}_{\Omega}$ the projectors to the
normalization region, which are known from (\ref{eq:Omegaprojection}).

\subsection{Weak coupling result}

A first limiting form of the QLBE that was obtained for the Maxwell-Boltzmann
distribution is the weak coupling result by one of us
{\cite{Vacchini2000a,Vacchini2001b,Vacchini2002a}} . Its derivation differs
strongly from the approach of the present article, using the van Hove
expression to relate the dynamic structure factor of the gas to the
differential cross section in the laboratory frame.

It can be regained from the present QLBE by replacing the exact scattering
amplitude $f$ in (\ref{eq:Ldef}) by its Born approximation $f_B$, which is
proportional to the Fourier transform of the interaction potential,
\begin{align}
  f_B \left( \tmmathbf{p}_f, \tmmathbf{p}_i \right)  = & - 4 \pi^2 \hbar
  m_{\ast} \langle \tmmathbf{p}_f |V \left( \mathsf{x} \right) |\tmmathbf{p}_i
  \rangle  \label{eq:fBorn}\\
   = & - \frac{m_{\ast}}{2 \pi \hbar^2} \int \mathd \tmmathbf{x} \, V \left(
  \tmmathbf{x} \right) \exp \left( - i \frac{\left( \tmmathbf{p}_f
  -\tmmathbf{p}_i \right) \cdot \tmmathbf{x}}{\hbar} \right) \bignone .
  \nonumber
\end{align}
Importantly, the Born approximation depends only on the momentum transfer
$\tmmathbf{p}_f -\tmmathbf{p}_i$, but not on the energy in the relative
motion. Even though $f_B$ violates the unitarity relation expressed by the
optical theorem, it can be used to approximate the scattering amplitude if the
energy of the relative motion is much larger than the interaction energy.

Inserting the Born approximation (\ref{eq:fBorn}) into the function
(\ref{eq:Ldef}) defining the jump operators $\mathsf{L}_{\tmmathbf{Q},
\tmmathbf{p}} = \mathe^{i \mathsf{X \cdot \tmmathbf{Q}/ \hbar}} L \left(
\tmmathbf{p}, \mathsf{P} ; \tmmathbf{Q} \right)$ from Sect.~\ref{sec:or}, one
notes that the $\tmmathbf{P}$-dependence {\tmem{drops out}} in the scattering
amplitude. As a result, the Born approximation of the function (\ref{eq:Ldef})
can be written as
\begin{eqnarray}
  L_B \left( \tmmathbf{p}, \tmmathbf{P}; \tmmathbf{Q} \right) & = & \left[
  \frac{n_{\tmop{gas}} }{m_{\ast}^2}  \right]^{1 / 2} f_{\text{B}} \left(
  -\tmmathbf{Q} \right) \sqrt{S \left( \tmmathbf{Q}, \tmmathbf{P} \right)}
  \nonumber\\
  &  & \times \frac{1}{\sqrt{\pi^{1 / 2} p_{\beta}}} \exp \left( -
  \frac{\tmmathbf{p}_{\bot \tmmathbf{Q}}^2}{2 p_{\beta}^2} \right), 
  \label{eq:LBdef}
\end{eqnarray}
where $S \left( \tmmathbf{Q}, \tmmathbf{P} \right)$ is the dynamic structure
factor of the Maxwell gas {\cite{Vacchini2001a}},
\begin{eqnarray}
  &  & S \left( \tmmathbf{Q}, \tmmathbf{P} \right) \nonumber\\
  &  & = \hspace{0.6em} \sqrt{\frac{\beta m}{2 \pi}} \frac{1}{Q} \exp \left(
  - \beta \frac{\left( \left( 1 + \frac{m}{M} \right) Q^2 + 2 \frac{m}{M}
  \tmmathbf{P} \cdot \tmmathbf{Q} \right)^2}{8 mQ^2} \right) . \nonumber\\
  &  &  \label{eq:dsf}
\end{eqnarray}
Since the $\tmmathbf{p}$-dependence in (\ref{eq:LBdef}) appears just as a
factor, one can carry out the $\tmmathbf{Q}^{\bot}$-integration in the
operator representation (\ref{eq:qlbe}) of the master equation. The weak
coupling approximation of the QLBE thus reduces to the form
\begin{eqnarray}
  \mathcal{L}^B \rho & = & \bigintlim \mathd \tmmathbf{Q} \left\{
  \widetilde{\mathsf{L}}_{\tmmathbf{Q}} \rho \widetilde{\mathsf{L}}
  _{\tmmathbf{Q}}^{\dag} - \frac{1}{2} \rho \widetilde{\mathsf{L}}
  _{\tmmathbf{Q}}^{\dag} \widetilde{\mathsf{L}} _{\tmmathbf{Q}} - \frac{1}{2} 
  \widetilde{\mathsf{L}} _{\tmmathbf{Q}}^{\dag} \widetilde{\mathsf{L}}
  _{\tmmathbf{Q}} \rho \right\} \bignone .  \label{eq:LcalB}
\end{eqnarray}
The corresponding jump operators
\begin{eqnarray}
  \widetilde{\mathsf{L}} _{\tmmathbf{Q}} & = & \mathe^{i \mathsf{X \cdot
  \tmmathbf{Q}/ \hbar}}  \left[ \frac{n_{\tmop{gas}} }{m_{\ast}^2} S \left(
  \tmmathbf{Q}, \mathsf{P} \right) \sigma_B \left( \tmmathbf{Q} \right)
  \right]^{1 / 2},  \label{eq:LQB}
\end{eqnarray}
are determined by the cross section in Born approximation, $\sigma_B \left(
\tmmathbf{Q} \right) = |f_{\text{B}} \left( \tmmathbf{Q} \right) |^2 =
|f_{\text{B}} \left( \um \tmmathbf{Q} \right) |^2$, rather than the individual
scattering amplitudes. As a result, the momentum operator $\mathsf{P}$ shows
up with a particularly simple functional dependence, given by the dynamic
structure factor (\ref{eq:dsf}). Recently, the behavior of this equation was
studied by means of a Monte Carlo simulation {\cite{Breuer2007a}}.

The weak coupling form of the QLBE coincides with the expression derived
earlier by one of us in Ref.~{\cite{Vacchini2001b}} (as can be seen if one
combines the Eqs.~(2) and (25) of that article, setting $\tilde{t} \left(
\tmmathbf{q} \right) = - f_B \left( \tmmathbf{q} \right) / 4 \pi^2 \hbar
m_{\ast}$ and $z = n \lambda_{\tmop{th}}^3$). This agreement is quite
remarkable, given the very different type of argumentation in
{\cite{Vacchini2001b}}, and it serves to corroborate the validity of the
present result.

Incidentally, Eq.~(\ref{eq:LcalB}) also shows that the full QLBE
{\tmem{cannot}} be obtained from the weak-coupling result by simply replacing
$f_B$ by the proper scattering amplitude. This procedure would be ambiguous
since the exact scattering amplitude is not just a function of the momentum
transfer. As discussed at the end of Sect.~\ref{sec:wp}, the dependence of the
scattering amplitudes on the tracer particle momentum, which dropped out in
the Born approximation, is required if one wants to cover the full set of
pairs of scattering trajectories allowed under energy and momentum
conservation.

In this context, it is worth mentioning that similar equations are obtained
from a heuristic method of dealing with products of delta functions like the
ones encountered in Sec.~\ref{sec:diagev}, see e.g.
{\cite{Diosi1995a,Adler2006a}}. There, one of the energy delta functions is
replaced by a finite Fourier integration over the `elapsed time' as is done in
derivations of Fermi's golden rule. Effectively, this amounts to a treatment
in second order perturbation theory where it is permissible to identify the
interaction Hamiltonian with the Born approximation of the T-matrix. Although
this brings scattering theory language into the game, one should not be
tempted to conclude that the non-perturbative equation can be obtained by
using the exact T-matrix.

\subsection{Index of refraction}\label{sec:ior}

An application of the QLBE involving a rather special limit concerns the
interference of matter waves in a Mach-Zehnder setup, where two interference
paths are spatially separated by a macroscopic distance. One may ask how the
interference fringe pattern changes if the particle is allowed to interact
with a background gas in one of the interferometer arms. This setup was
realized experimentally with Na and Li atoms
{\cite{Schmiedmayer1995a,Vigue1995a,Jacquey2007a}} (and it is a common
configuration in neutron interferometry where the ``background gas'' consists
rather of thermalized condensed matter {\cite{Sears1989a,Werner2000a}}).

In these situations the beam is strongly collimated, while the likelihood of
double collisions is small, so that after any collision that changes the
momentum of the interfering tracer particle the latter will be blocked by the
interferometer apertures. As a consequence, only forward-scattered amplitudes
may contribute to the interference pattern, thus making the energy shift
(\ref{eq:cohexp2}) directly observable as a change in the phase of the
interference pattern. At the same time an attenuation of the recorded signal
is observed.

The phase shift is usually accounted for by attributing a real index of
refraction $n_1$ to the gas, which describes the modification of the de
Broglie wavelength due to the energy shift. Exploiting the analogy between the
force-free Schr\"odinger equation and the Helmholtz wave equation
{\cite{Sears1989a,Adams1994a}} the index of refraction for matter waves is
determined by the ratio of the energy shift $E_{\text{n}} \left( \tmmathbf{P}
\right)$ from (\ref{eq:cohexp2}) to the vacuum kinetic energy $E_{\tmop{kin}}
= P^2 / 2 M$ of the particle,
\begin{eqnarray}
  n_1^2 & = & 1 - \frac{E_{\text{n}} \left( \tmmathbf{P}
  \right)}{E_{\tmop{kin}} \left( \tmmathbf{P} \right)}  \label{eq:EnoEkin}\\
  & = & 1 + 4 \pi \hbar^2 \frac{n_{\tmop{gas}}}{P^2} \frac{M}{m_{\ast}}  \int
  \mathd \tmmathbf{p}_0 \, \mu \left( \tmmathbf{p}_0 \right) \nonumber\\
  &  & \times \tmop{Re} \left[ f \left( 0 ; \frac{1}{2 m_{\ast}} \left[
  \tmop{rel} \left( \tmmathbf{p}_0, \tmmathbf{P} \right) \right]^2 \right)
  \right] . \nonumber
\end{eqnarray}
Here we took a rotationally invariant scattering amplitude, $f \left(
\tmmathbf{p}_f, \tmmathbf{p}_i \right) = f \left( \theta ; E_{\tmop{rel}}
\right)$, with $\theta = \sphericalangle {\left( \tmmathbf{p}_f,
\tmmathbf{p}_i \right)}$ and $E_{\tmop{rel}} = p_i^2 / 2 m_{\ast}$.

The index of refraction is typically close to unity, and therefore well
approximated by the linearization
\begin{eqnarray}
  n_1 & = & 1 + 2 \pi \frac{n_{\tmop{gas}}}{K^2}  \frac{M}{m_{\ast}} \tmop{Re}
  \langle f \rangle,  \label{eq:n1}
\end{eqnarray}
where $K = P / \hbar$ is the wavenumber of the interfering particle and
$\tmop{Re} \langle f \rangle$ denotes the real part of the thermally averaged
forward scattering amplitude,
\begin{equation}
  \langle f \rangle  = \int \mathd \tmmathbf{p}_0 \, \mu \left(
  \tmmathbf{p}_0 \right) f \left( 0 ; \frac{1}{2 m_{\ast}} \left[ \tmop{rel}
  \left( \tmmathbf{p}_0, \tmmathbf{P} \right) \right]^2 \right) . 
  \label{eq:fav}
\end{equation}

It is common in optics to account for the absorption in a medium by
introducing an imaginary part to the index of refraction, which describes the
exponential decay of the beam intensity. In the case of a background gas the
tracer particles do not get absorbed, of course. However, for a strongly
collimated particle beam one expects an exponential decay of the beam
intensity after a distance $L$, since collisions with the background gas
decrease the probability of remaining in the beam, thus reducing the fraction
of particles taking part in the coherent, wavelike behavior. The decay may be
described by neglecting the gain term in Eq.~(\ref{eq:QLBEcl}), and
integrating the remaining equation $\partial^{\tmop{coll}}_t f_{\text{p}}
(\tmmathbf{P}) = - M_{\tmop{out}}^{\tmop{cl}} \left( \tmmathbf{P} \right)
f_{\text{p}} (\tmmathbf{P})$ up to time $t = L / V$, with $V = P / M$, yields
the reduction factor $\exp \left( - M_{\tmop{cl}}^{\tmop{out}} \left( P
\right) L / V \right)$. By comparing this to the damped intensity of a wave,
$\exp \left( - 2 n_2 KL \right)$, one finds
\begin{eqnarray}
  n_2 & = & \frac{M_{\tmop{cl}}^{\tmop{out}} \left( \hbar K \right)}{2 \hbar
  K^2 / M} . 
\end{eqnarray}
Inserting $M_{\tmop{cl}}^{\tmop{out}}$ from (\ref{eq:Mcloutexp2}) and using
the optical theorem {\cite{Taylor1972a}}, that is, $p_i \sigma_{\tmop{tot}}
\left( \tmmathbf{p}_i \right) = 4 \pi \hbar \tmop{Im} \left[ f \left(
\tmmathbf{p}_i, \tmmathbf{p}_i \right) \right]$, one gets an expression
analogous to (\ref{eq:n1}),
\begin{eqnarray}
  n_2 & = & 2 \pi \frac{n_{\tmop{gas}}}{K^2}  \frac{M}{m_{\ast}} \tmop{Im}
  \langle f \rangle, 
\end{eqnarray}
with $\tmop{Im} \langle f \rangle$ the imaginary part of (\ref{eq:fav}). It
follows that the combined effect of the energy shift and the reduction of the
amplitude of the coherent beam can be described by a complex index of
refraction $n = n_1 + in_2 = 2 \pi n_{\tmop{gas}} M / \left( m_{\ast} K^2
\right) \langle f \rangle$.

In case of a Maxwell-Boltzmann distribution (\ref{eq:muMB}) the average takes
the form
\begin{eqnarray}
  \langle f \rangle & = & \frac{2}{\sqrt{\pi^{}}} \int_0^{\infty} \frac{\mathd
  v}{v_{\beta}}  \frac{v}{V} \sinh \left( \frac{2 vV}{v_{\beta}^2} \right)
  \exp \left( - \frac{v^2 + V^2}{v_{\beta}^2} \right)  \nonumber\\
  &  & \times f \left( 0 ; \frac{m_{\ast}}{2} v^2 \right), 
\end{eqnarray}
with $V = \hbar K / M$ the velocity of the interfering particle and $v_{\beta}
= p_{\beta} / m$. This expression of the thermally averaged forward scattering
amplitude coincides with the one obtained by C. Champenois and collaborators
in Ref.~{\cite{Champenois1999a,Champenois2007a}} with a very different
argumentation. It is used in the analysis of the recent experiment with Li
atoms {\cite{Jacquey2007a}}. We note that the earlier experiments
{\cite{Schmiedmayer1995a,Vigue1995a}} and the corresponding theoretical
treatments {\cite{Forrey1996a,Kharchenko2001a}} were based on different
expressions for $\langle f \rangle$ which we consider incorrect, see also the
discussion in {\cite{Champenois2007a}}.

\subsection{Diffusive limit}

A final important border case is the diffusive limit which is applicable if
the tracer state is close to thermal and if its mass is much greater than the
gas particle mass, so that the motion is characterized by small momentum
transfers. As discussed in {\cite{Vacchini2007a}}, an expansion of the jump
operators (\ref{eq:Lopdef}) to second order in the tracer position and
momentum operators $\mathsf{X}$ and $\mathsf{P}$ is then permissible. In the
special case of a {\tmem{constant}} scattering cross-section $\left| f \left(
\tmmathbf{p}_f, \tmmathbf{p}_i \right) \right|^2 =
\sigma_{\tmop{tot}}^{\text{c} \tmop{onst}} / 4 \pi$ the QLBE then transforms
into the generalized Caldeira-Leggett master equation
\begin{eqnarray}
  \tilde{\mathcal{L}} \rho & = & - \frac{i}{\hbar} \frac{\eta}{2} \sum^3_{i =
  1} \left[ \mathsf{X}_i, \left\{ \mathsf{P}_i, \rho \right\} \right] -
  \frac{D_{pp}}{\hbar^2} \sum^3_{i = 1} \left[ \mathsf{X}_i, \left[
  \mathsf{X}_i, \rho \right] \right] \nonumber\\
  &  & - \frac{D_{xx}}{\hbar^2} \sum^3_{i = 1} \left[ \mathsf{P}_i, \left[
  \mathsf{P}_i, \rho \right] \right] .  \label{eq:CLext}
\end{eqnarray}
It differs form the original equation {\cite{Caldeira1983a}} in the presence
of the last term on the r.h.s of (\ref{eq:CLext}), which is necessary to
ensure the complete positivity of the dynamics generated by
$\widetilde{\mathcal{L}}$ {\cite{Breuer2007b}}. We emphasize that, unlike in
derivations using phenomenological choices for the model environment
{\cite{Haake1985a,Grabert1988a,Unruh1989a,Hu1992a}}, the friction and
diffusion coefficients $\eta$ and $D_{pp}$ are here uniquely specified by
physically measurable properties of the gas. Specifically, the calculation in
{\cite{Vacchini2007a}} shows that the friction coefficient $\eta$ is
determined by the temperature, the mass, and the density of the gas, as well
as by the scattering cross section,
\begin{eqnarray}
  \eta & = & \frac{8}{3 \pi^{1 / 2}} \frac{n_{\tmop{gas}} p_{\beta}
  \sigma^{\tmop{const}}_{\tmop{tot}}}{M} .  \label{eq:eta}
\end{eqnarray}
The momentum diffusion constant $D_{pp}$ is related to $\eta$ by the
fluctuation-dissipation relation
\begin{eqnarray}
  D_{pp} & = & \frac{\eta M}{\beta} .  \label{eq:Dppfinal}
\end{eqnarray}
Moreover, the coefficient of the ``position-diffusion'' term $D_{xx}$, is
already determined by $\eta$ and $D_{pp}$, and it is given by the smallest
value compatible with complete positivity{\footnote{Di\'osi's equation
{\cite{Diosi1995a}} leads to the same structure (\ref{eq:CLext}), but the
``position diffusion'' constant $D_{x x}$ is a complicated function of the
cross section instead of being simply related to $D_{p p}$ and $\beta .$}}
{\cite{Breuer2007b}},
\begin{eqnarray}
  D_{xx} & = & \eta \frac{\hbar^2 \beta}{16 M} \hspace{0.6em} = \hspace{0.6em}
  \left( \frac{\hbar^{} \beta}{4 M} \right)^2 D_{pp} .  \label{eq:Dxxfinal}
\end{eqnarray}
This shows that the diffusive limit turns the QLBE into the closest possible
quantum analogue to the corresponding classical Kramers equation
{\cite{Risken1989a}}.

\section{Conclusions}\label{sec:cc}

As seen in the previous section, all relevant limiting cases of the QLBE lead
naturally to established master equations. In conjunction with the detailed
derivations presented in Sects.~\ref{sec:momentum}--\ref{sec:refrac}, this
provides ample evidence that Eq.~(\ref{eq:qlbe2}) is the appropriate full
quantum analogue of the classical linear Boltzmann equation. As such, it
serves to describe non-perturbatively and in a unified framework the effects
of decoherence and dissipation on a tracer particle.

One reason that seems to have prevented this equation from being formulated
earlier is the curious appearance of a second momentum integral in
(\ref{eq:qlbe}) which, in addition to the integration over the momentum
exchange $\tmmathbf{Q}$, runs over the plane perpendicular to $\tmmathbf{Q}$.
This makes the equation a bit cumbersome at first sight, at least if
represented in a specific basis. However, we have seen in the course of the
derivation that this five-dimensional integration is necessary if one wants to
cover all the pairs of scattering trajectories which are allowed by both the
energy and momentum conservation and by the choice of $\tmmathbf{Q}$. From a
quantum mechanical point of view it is indeed natural to expect that the full
set of possible scattering amplitudes contributes to the dynamics. The
somewhat unwieldy explicit form is then the inevitable result of the
transformation from the center of mass frame, where the scattering
transformation takes place, to the laboratory frame needed for the tracer
particle coordinate.

Needless to say, the QLBE has a number of limitations. Like the classical
linear Boltzmann equation, it cannot be applied in environments where the
central Markov assumption is inappropriate, such as liquids. Moreover, it is
not applicable at temperatures where the gas is quantum degenerate, and it is
far from obvious how this possibility could be incorporated in the framework
of the monitoring approach. Finally, let us reiterate that we presented here a
physical derivation which, though stringent and leading to a uniquely
distinguished equation, may be hard to substantiate from a formal point of
view. An alternative, mathematically more rigorous derivation would be
certainly desirable.

{\subsubsection*{Acknowledgements}}

We thank J.~Vigu\'e for helpful discussions on the index of refraction for
matter waves. The work was partially supported by the DFG Emmy Noether program
(KH), and by the Italian MUR under PRIN2005 (BV).

\end{document}